\documentclass[twocolumn,prd,nofootinbib,preprintnumbers,superscriptaddress]{revtex4}

\pdfoutput=1
\usepackage{graphicx}
\usepackage{amsmath,bm}
\usepackage[T1]{fontenc}
\usepackage{rotating}
\usepackage{float}
\usepackage{aas_macros}
\usepackage{booktabs}
\floatstyle{plaintop}
\usepackage{mathtools} 
\usepackage{braket}
\usepackage{tikz}

\usepackage{color}
\definecolor{naviBlue}{RGB}{0,0,128}
\usepackage[colorlinks  = true,
            linkcolor   = naviBlue,
            urlcolor    = naviBlue,
            citecolor   = naviBlue,
            anchorcolor = naviBlue]{hyperref}
\newcommand{\secref}[1]{\hyperref[sec::#1]{SECTION~\ref*{sec::#1}}}
\newcommand{\subsecref}[1]{\hyperref[subsec::#1]{SECTION.~\ref*{subsec::#1}}}
\newcommand{\figref}[1]{\hyperref[fig::#1]{FIG.$\,$\ref*{fig::#1}}}
\newcommand{\tabref}[1]{\hyperref[tab::#1]{TABLE$\,$\ref*{tab::#1}}}
\newcommand{\eqnref}[1]{\hyperref[eqn::#1]{Eq.$\,$(\ref*{eqn::#1})}}

\newcommand{\diff}{\mathrm{d}}

\newcommand{\eq}{\mathrm{eq}}

\def\beq{\begin{equation}}
\def\eeq{\end{equation}}
\definecolor{darkgreen}{RGB}{0,170,0}
\definecolor{darkgray}{RGB}{110,110,108}
\newcommand{\bea}{\begin{eqnarray}}
\newcommand{\eea}{\end{eqnarray}}
\newcommand{\be}{\begin{equation}}
\newcommand{\ee}{\end{equation}}
\newcommand{\nn}{\nonumber}

\usepackage{array}
\newcolumntype{C}{>{$}c<{$}} 	%use math mode in table column by default

\definecolor{purple}{RGB}{160,0,160}
\definecolor{plotpink}{RGB}{205,0,180}
\definecolor{plotcyan}{RGB}{0,215,215}
\definecolor{darkcyan}{RGB}{17,155,155}

\definecolor{plotblue}{RGB}{0,0,235}
\definecolor{plotorange}{RGB}{245,140,0}
\definecolor{darkorange}{RGB}{210,100,0}
\definecolor{plotgreen}{RGB}{30,130,0}
\definecolor{plotred}{RGB}{240,0,0}
\definecolor{darkred}{RGB}{180,0,0}
\definecolor{darkergreen}{RGB}{0,130,3}
\definecolor{darkgreen}{RGB}{0,160,0}
\definecolor{lblue}{RGB}{100,130,205}
\definecolor{ourbrown}{RGB}{151,105,56}
\definecolor{darkblue}{RGB}{10,10,145}
\definecolor{gray}{RGB}{90,90,90}
\definecolor{graycyan}{RGB}{90,113,113}
\definecolor{dgraycyan}{RGB}{90,113,113}
\definecolor{darkgraycyan}{RGB}{66,84,84}
\definecolor{darkH}{RGB}{11,65,188}

\newcommand{\avb}[1]{\big\langle #1 \big\rangle}

\def\beq{\begin{equation}}
\def\eeq{\end{equation}}

\newcommand{\E}{\mathrm{e}}

\newcommand{\ie}{\emph{i.e.}}
\newcommand{\eg}{\emph{e.g.}}
\newcommand{\cf}{\emph{cf.}}

\newcommand{\zbt}{\tilde{\zeta_b}}

\newcommand{\zb}{\zeta_b}
\newcommand{\zs}{\zeta_s}
\newcommand{\ellp}{\ell^{\prime}}
\newcommand{\np}{n^{\prime}}
\newcommand{\kt}{\tilde{\kappa}}
\newcommand{\kp}{\kappa^\prime}
\newcommand{\kpt}{\tilde{\kappa}^{\prime}}

\begin{document}

\title{
Excited bound states and their role in dark matter production
}

\author{Tobias Binder}
\email{tobias.binder@tum.de}
\affiliation{Physik Department T31, Technische Universit\"at M\"unchen,
James-Franck-Stra\ss e 1, D-85748 Garching, Germany}

\author{Mathias Garny}
\email{mathias.garny@tum.de}
\affiliation{Physik Department T31, Technische Universit\"at M\"unchen,
James-Franck-Stra\ss e 1, D-85748 Garching, Germany} 

\author{Jan Heisig}
\email{heisig@physik.rwth-aachen.de}
\affiliation{Institute for Theoretical Particle Physics and Cosmology, RWTH Aachen University, D-52056 Aachen, Germany}
\affiliation{Department of Physics, University of Virginia, Charlottesville, Virginia 22904-4714, USA}

\author{Stefan Lederer}
\email{stefan.lederer@tum.de}
\affiliation{Physik Department T31, Technische Universit\"at M\"unchen,
James-Franck-Stra\ss e 1, D-85748 Garching, Germany}

\author{Kai Urban}
\email{kai.urban@tum.de}
\affiliation{Physik Department T31, Technische Universit\"at M\"unchen,
James-Franck-Stra\ss e 1, D-85748 Garching, Germany}

\preprint{TUM-HEP 1469/23}
\preprint{TTK-23-21}

\begin{abstract}
We explore the impact of highly excited bound states on the evolution of number densities of new physics particles, specifically dark matter, in the early Universe. Focusing on dipole transitions within perturbative, unbroken gauge theories, we develop an efficient method  for including around a million bound state formation and bound-to-bound transition processes. This enables us to examine partial-wave unitarity and accurately describe the  freeze-out dynamics down to very low temperatures. In the non-Abelian case, we find that highly excited states can prevent the particles from freezing out, supporting a continuous depletion in the regime consistent with perturbativity and unitarity. We apply our formalism to a simplified dark matter model featuring a colored and electrically charged $t$-channel mediator. Our focus is on the regime of superWIMP production which is commonly characterized by a mediator freeze-out followed by its late decay into dark matter. In contrast, we find that excited states render mediator depletion efficient all the way until its decay, introducing a dependence of the dark matter density on the mediator lifetime as a novel feature. The impact of bound states on the viable dark matter mass can amount to an order of magnitude, relaxing constraints from Lyman-$\alpha$ observations.
\end{abstract}

\maketitle
\tableofcontents

%===================================================================
\section{Introduction}\label{sec:intro}
%===================================================================

Understanding  the composition of matter in our Universe constitutes a major challenge of today's fundamental physics. Notably, explanations of both the observed dark matter density and matter-antimatter asymmetry necessitate the introduction of physics beyond the Standard Model (SM) and therewith the computation of interactions among new -- and presumably heavy -- particles in the early Universe. If such new particles interact via a light force carrier, a significant contribution to their depletion may be given by the formation and subsequent decay of bound states, which has intriguing consequences for their thermal history. For instance, in the context of electroweakly charged dark matter~\cite{Mitridate:2017izz, Bottaro:2021snn, Bottaro:2022one} and colored coannihilation scenarios~\cite{Ellis:2015vaa,Ellis:2015vna,Liew:2016hqo,Kim:2016zyy,Harz:2017dlj,Harz:2018csl,Biondini:2018pwp,Biondini:2018ovz,Fukuda:2018ufg,Biondini:2019int,Garny:2021qsr,Becker:2022iso}, it has been shown that the inclusion of bound state effects can strongly alter the prediction for the relic density.

Generally, we can classify radiative bound state formation (BSF) processes in terms of its leading multipole contribution:

\begin{enumerate}
    \item \emph{Monopole:} Bound-state formation via emission of a \emph{charged scalar field} can be extremely relevant~\cite{Oncala:2019yvj,Oncala:2021swy}. As the emission carries away charge, it changes the initial and final two-particle state, leading to non-orthogonal states and ultimately to a non-vanishing monopole contribution. The BSF cross section via monopole transitions has been worked out for arbitrary excited states. However, partial-wave unitarity can be problematic already for capture into the ground state~\cite{Oncala:2019yvj}. 
    \item \emph{Dipole:} Known examples where radiative BSF is dominated by the contribution of the (dark electric) dipole moment, are  $U(1)$~\cite{vonHarling:2014kha,Kamada:2019jch, Binder:2020efn, Biondini:2023zcz} or $SU(N_c)$~\cite{Harz:2018csl,DeLuca:2018mzn, Asadi:2021pwo, Binder:2021otw, Asadi:2022vkc, Becker:2023opo, Assi:2023cfo, Biondini:2023zcz} gauge symmetry extensions of the SM\@. In these cases, the emitted particle is a massless gauge \emph{vector field}. Another possibility considers the accompanying dark matter particle to have SM electroweak~\cite{Asadi:2016ybp,Mitridate:2017izz, Bottaro:2021snn, Bottaro:2022one} and/or color charge~\cite{Ellis:2015vaa,Ellis:2015vna,Liew:2016hqo,Kim:2016zyy,Harz:2017dlj,Harz:2018csl,Biondini:2018pwp,Biondini:2018ovz,Fukuda:2018ufg,Biondini:2019int,Garny:2021qsr,Becker:2022iso,Biondini:2022ggt}. A famous example of the latter is squark %stop
    coannihilation in the context of the Minimal Supersymmetric extension of the Standard Model (MSSM), or simplified $t$-channel mediator models inspired by it. 
    \item  \emph{Quadrupole:} One example where quadrupole moments contribute at the leading order, considers the emission of a \emph{real scalar field} in the radiative bound state formation process~\cite{Wise:2014jva,Petraki:2015hla,An:2016kie,Biondini:2021ycj,Biondini:2021ccr}. Beyond capture into the ground state, little is known about higher excited states and bound-to-bound transitions.
\end{enumerate}
In this work, we will present a more detailed investigation of the second case, where BSF and transitions among bound states in unbroken gauge theories are dominated by the (chromo) electric dipole contribution. While it is indeed the most considered scenario, it still remains unclear by how much the inclusion of highly excited bound states into the chemical network contributes to the depletion of the dark matter relic density. As shown recently in Ref.~\cite{Garny:2021qsr}, the impact of higher excitations can be sizeable, in particular, when considering scenarios beyond the paradigm of weakly interacting massive particles (WIMPs), such as conversion-driven freeze-out~\cite{Garny:2017rxs}, and for bound states driven by a perturbative, unbroken non-Abelian gauge symmetry. 

The greatest obstacle is the accurate evaluation of the dipole matrix elements for BSF and for transitions among highly excited states, especially in non-Abelian theories. General formulas to evaluate the dipole matrix elements for all principle and angular momentum quantum numbers, $n$ and $\ell$, respectively, have been provided in~\cite{Garny:2021qsr}. Here, we significantly improve on efficiency and numerical stability of their evaluation, which allows us to explore the contribution of up to half a million bound states (all $n\leq1000,\; \ell\leq n-1$ states) when considering BSF\@. For bound-to-bound matrix elements, we are able to include all electric dipole transitions up to $n\leq100,\; \ell\leq n-1$, which are about one million transitions in total for the processes allowed by the selection rules.  

These improvements allow us to address several key scientific questions. First, we consider the velocity dependence of the BSF cross section in vacuum and investigate partial wave unitary properties in Abelian and non-Abelian gauge theories. For the case of $SU(N_c)$, the inclusion of a large number of excitations sheds light on the breakdown of our theoretical framework and the necessity of its unitarization. 

Next, we consider the interplay of BSF, ionization, bound-to-bound transitions and bound state decays in the thermal bath  of  the early Universe. Following the formalism of~\cite{Garny:2021qsr,Binder:2021vfo}, we describe their effect on the bound state constituents' abundance via an effective thermally averaged cross section. Focusing on the perturbative coupling regime which turns out to be consistent with unitarity, we pose the important question whether the effective cross section grows slower or faster than the inverse temperature, implying freeze-out or a continuous depletion of the abundance, respectively. The latter case is found for non-Abelian interactions and yields important phenomenological implications.

We exemplify these implications in detail in the last part of our work, where we apply our numerical framework to the superWIMP scenario~\cite{Covi:1999ty,Feng:2003uy}, considering a simplified model with a colored and electrically charged $t$-channel mediator~\cite{Garny:2018ali,Decant:2021mhj}.
We showcase the  effect of highly excited bound states and find that the combined effect of strong and elecromagnetic interactions conspire to reduce the relic density by more than an order of magnitude compared to the case when including Sommerfeld-enhanced annihilations only. Thereby we improve on earlier results within this scenario considering the ground state only~\cite{Decant:2021mhj,Bollig:2021psb}. We find that this has important implications on the viable parameter space, in particular to relief Lyman-$\alpha$ constraints.

The remainder of this paper is organized as follows.
In Sec.\,\ref{sec:VelocityDependence}, we discuss BSF in vacuum and investigate the velocity-dependence of the BSF cross section in view of partial wave unitarity. In Sec.\,\ref{sec:darksectors}, we study 
the scaling of the effective cross section with temperature and discuss the implications for freeze-out. The setup and relevant quantities for the $t$-channel model are reviewed in Sec.\,\ref{sec:toymodel}, and our results for the impact of highly excited states on the superWIMP mechanism for dark matter production are discussed in Sec.\,\ref{sec:swimp}\@. We conclude in Sec.\,\ref{sec:conclusion}. Appendix~\ref{sec:bsfapp} details the evaluation of BSF and transition matrix elements and App.\,\ref{sec:AppRates} provides expressions for the cross sections and rates used in our analysis. Finally, in App.\,\ref{sec:uniQED}, we show implications for the cosmologically viable parameter space of dark QED including highly excited bound states.

%===================================================================
\section{Bound-state formation in vacuum}\label{sec:VelocityDependence}
%===================================================================

%-------------------------------------------------------------------
\subsection{Matrix elements}
%-------------------------------------------------------------------

Our starting point for computing radiative BSF in gauge theories is \emph{potential non-relativistic effective field theory}~\cite{Pineda:1997bj, Beneke:1998jj, Beneke:1999zr,Brambilla:1999xf}. In this framework, the interaction of two non-relativistic particles with $SU(N_c)$ or $U(1)$ gauge vector fields at the ultra-soft scale can effectively be described by a \emph{(chromo-)electric dipole operator}, $g\, \mathbf{r} \cdot \mathbf{E}$, with gauge coupling $g$, relative distance $\mathbf{r}$, and electric field $\mathbf{E}$. In the two-particle subspace of the non-relativistic particles, this operator leads to matrix elements of the form
\begin{align}
\bra{\psi_f} \mathbf{r} \ket{\psi_i} = \int \text{d}^3 r \; \psi_f^\star(\mathbf{r}) \; \mathbf{r} \; \psi_i(\mathbf{r}).\label{eq:dipolmatrix}
\end{align}
The squared absolute value of these matrix elements is directly related to our physical quantities of interest: bound-state formation cross sections and bound-to-bound rates for electric dipole transitions. For concrete examples related to dark matter, see Refs.~\cite{Binder:2020efn,Binder:2021otw,Biondini:2023zcz,Biondini:2023yxt}.

In App.~\ref{sec:bsfapp}, we develop an efficient way of evaluating the matrix elements for systems, where the initial state $\psi_i$ and the final state $\psi_f$ are the solutions of two-body Schr\"odinger equations with corresponding potentials of Coulomb type:
\begin{align}
V_{i,f}(r)= -\frac{\alpha_{i,f}^\text{eff}}{r}\;.
\end{align}
The effective coupling strength $\alpha^\text{eff}$ of the initial and final state incorporates the details of the underlying particle physics model. We will consider Abelian gauge theories where the effective couplings are equal, and non-Abelian gauge theories where they can be different. We denote the effective couplings by $\alpha_b^\text{eff}$ and $\alpha_s^\text{eff}$ when referring to bound and scattering states, respectively.
In the following, we explore the contribution of highly excited bound states in concrete realisations.

%-------------------------------------------------------------------
\subsection{Abelian case}
%-------------------------------------------------------------------

As a first concrete model, we consider Quantum Electrodynamics (QED) in the non-relativistic regime of the Fermionic particles. The two-particle states of interest then consist of two oppositely charged particles forming gauge singlets. Standard Model examples are hydrogen recombination and positronium formation. In QED, the potential of both the initial and final state is attractive with identical strength. The corresponding BSF cross section, describing the electric dipole transition process of a scattering state into a bound state with quantum numbers $n$ and $\ell$, is
\begin{align}\label{eq:sigmaBSFdef}
(\sigma v)_{n\ell}=\frac{4 \alpha}{3} \Delta E^3 |\bra{\psi_{n\ell}} \mathbf{r} \ket{\psi_{\mathbf{p}}}|^2,
\end{align}
where $\alpha$ is the fine-structure constant. The difference of the initial and final state energy is the positive quantity $\Delta E = \frac{\mathbf{p}^2}{2 \mu} + E_{{\cal B}_{n\ell}}$, where $\mathbf{p}^2=\mu^2 v^2$ with $v$ being the relative velocity and $\mu$ the reduced mass. Here, the absolute value of the binding energy is given by
$E_{{\cal B}_{n\ell}}=\frac{\mu \alpha^2}{2 n^2}$ as in QED $\alpha=\alpha_i^\text{eff}=\alpha_f^\text{eff}$. The BSF cross section as defined in Eq.~\eqref{eq:sigmaBSFdef} is averaged over initial and summed over final spin degrees of freedom, as well as summed over the magnetic quantum numbers of the bound state (see App.~\ref{sec:AppRates} for details). Since the electric dipole operator is spin conserving, the same equation that applies to the Fermionic case (QED) also applies to, \eg, a complex scalar field charged under a $U(1)$ gauge symmetry~\cite{Petraki:2015hla}. For simplicity, we will commonly refer to both cases as $U(1)$ in the following, as we are mainly interested in physics beyond the SM and we would like to cover both the Fermionic and complex scalar case simultaneously in our discussion.

\begin{figure}
    \centering
    \includegraphics[scale=0.73]{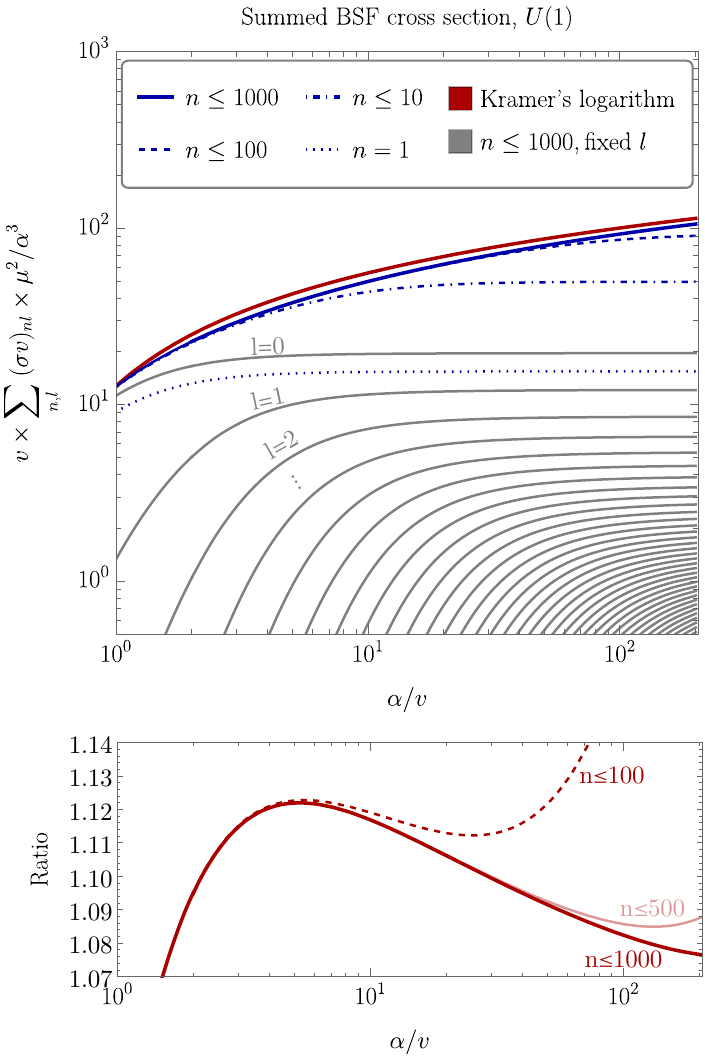}
    \caption{\emph{Upper:} Bound-state formation cross section  Eq.~(\ref{eq:sigmaBSFdef}) for $U(1)$, summed over all possible final bound-state quantum numbers $n$ and $\ell \leq n-1$. From the truncation of the sum for $n\leq1,10,100,1000$ (blue), one can infer that higher excited states contribute more for decreasing velocity. 
    The red line shows the analytic approximation Eq.~\eqref{eq:Kramer} when summing over all $n,\,\ell$, 
    known as Kramer`s logarithm. The gray lines show the contribution to the sum for fixed $\ell$, all of which approach a constant value at large $\alpha/v$, respecting partial-wave unitarity. \emph{Lower:} Ratio of Eq.\,\eqref{eq:Kramer} and the summed result.}
    \label{fig:svQED}
\end{figure}

We numerically evaluate the $U(1)$ BSF cross section in Eq.~\eqref{eq:sigmaBSFdef}, as detailed in App.~\ref{sec:bsfapp}, for various $n,\ell$. We present the result in Fig.~\ref{fig:svQED} in a model independent way, \ie~we multiply the cross section by $v \mu^2/\alpha^3$ to (i) show the remaining dependence on $\alpha/v$ and (ii) to highlight deviations of the velocity dependence from $1/v$. Let us begin with the well-known case of capture into the ground state $n=1$, $\ell=0$ (blue dotted line). For $v \ll \alpha$, the cross section for capture into the ground state scales as $1/v$ so that the shown product, $v \times (\sigma v)$, approaches a constant (see \eg~Ref.~\cite{Petraki:2015hla}, also regarding its magnitude relative to annihilation).  Similarly, we find the same velocity scaling when fixing the final state $\ell$ and summing over all $n \leq 1000$ (gray lines). Specifically, the $\ell=0$ gray line is larger by a factor 1.268
 than the $n=1$, $\ell=0$ blue dotted line (capture into the ground state). We note that this factor is smaller than the upper bound 1.6 derived analytically in Refs.~\cite{vonHarling:2014kha,Bethe:1957ncq}. For $\ell=0$, we additionally checked that our result for each $n\leq5$ coincides with the analytic results available in Ref.~\cite{Petraki:2016cnz} and up to $n=10$ for all $\ell$ with those of \cite{Garny:2021qsr}.   

Interestingly, when summing the BSF cross section both in $n$ and $\ell$ (blue lines), the velocity scaling becomes stronger than $1/v$. Here, we sum all $n\leq 10$ (dot-dashed), $100$ (dashed), $1000$ (solid) and always $\ell\leq n-1$ accordingly. We compare our summed result to the well-known Kramer's logarithm~\cite{Kramers1923XCIIIOT,1978JETP...48..639K} (red line):
\begin{align}\label{eq:Kramer}
\sum_{n,\ell}(\sigma v)_{n\ell} \simeq \frac{32 \pi}{3 \sqrt{3}} \frac{\alpha^2}{\mu^2} \frac{\alpha}{v}[\log(\alpha/v) + \gamma_E] \, \text{, for } v \ll \alpha.
\end{align}
The ratio of the Kramer's logarithm and our fully summed numerical result is shown in the bottom panel of Fig.~\ref{fig:svQED}. For $v \ll \alpha$ this ratio is expected to approach unity when including a very large number of excited states. We confirm this trend within the range of our numerical limitations, $n\leq1000$, $\ell\leq n-1$, which can also be seen as a non-trivial check of our code.\footnote{For a given $\alpha/v$, the amount of excited states needed for a convergent sum can be estimated. To this end, we consider the sum that leads to the Kramer's logarithm:
\begin{align}
\sum_n \frac{1}{n[n^2+(\alpha/v)^2]} \simeq [\log(\alpha/v) + \gamma_E] \text{, for } v \ll \alpha\,.
\end{align}
From the denominator, one can estimate that for a percentage accuracy, the maximum principle quantum number $n$ needs to be roughly an order of magnitude larger than a given $\alpha/v$. This is also what we observe for our summed numerical result in Fig.~\ref{fig:svQED}. For instance, summing all bound state contributions up to our numerical limit $n\leq1000,\ell\leq n-1$, provides a percentage-level accuracy for $\alpha/v \lesssim 100$ only, while $n\leq100,\ell\leq n-1$ would require  $\alpha/v \lesssim 10$ for the same accuracy.} 
The Kramer's logarithm has also been mentioned in earlier dark matter related works~\cite{Belotsky:2015osa,  An:2016gad,Petraki:2016cnz,Smirnov:2019ngs}.

Although the logarithm leads to a slope steeper than $1/v$, partial-wave unitarity is not violated here as the sum over different $\ell$ automatically includes different initial state angular momenta. However, we have checked that each individual angular momentum contribution of the initial state does not violate partial-wave unitarity as it \emph{does} scale as $1/v$ for $v \ll \alpha$. For instance, in this limit, the BSF cross section of the $\mathrm{s} \to n \mathrm{p}$ processes summed over all $2 \leq n\leq1000$ is larger by a constant factor $3.8$ than the $\rm{s}\to 2\rm{p}$.  Each angular momentum contribution therefore remains below the partial wave unitarity limit for all $v$, provided the coupling is sufficiently small.
In the non-Abelian case, we will observe a qualitatively different behavior, that implies partial wave unitarity violation even for (in principle) arbitrarily small couplings.

%-------------------------------------------------------------------
\subsection{Non-Abelian case}
%-------------------------------------------------------------------

As our second example, we consider a non-Abelian $SU(N_c)$ gauge theory, specifically $SU(3)$. Interestingly, it provides a qualitatively different phenomenology from QED even though in both cases the leading BSF processes are based on dipole transitions. While in QED the initial and final state potentials are both \emph{attractive} with the same strength, in Quantum Chromodynamics (QCD), the initial state potential of the adjoint pair is \emph{repulsive}. As we shall point out and explore in the following, this feature is accompanied by partial-wave unitarity violation in QCD or in a general $SU(N_c)$ for $N_c \geq 2$. 

In particular, we consider pairs of non-relativistic Fermionic particles in the fundamental and anti-fundamental representation of $SU(3)$, \ie~the two-particle space is spanned by the direct sum of singlet and adjoint pair states:  $ 3 \otimes \bar{3} =1 \oplus 8 $. A SM example is heavy quarkonium formation in SM QCD\@. The BSF cross section describing the chromo electric dipole transition process of an adjoint scattering state into a singlet bound state is given by
\begin{align}
(\sigma v)_{n\ell}= \frac{C_F}{N_c^2}\frac{4 \alpha}{3} \Delta E^3 |\bra{\psi_{n\ell}^{[{\bf1}]}} \mathbf{r} \ket{\psi_{\mathbf{p}}^{[{\bf adj}]}}|^2,\label{eq:sigmaBSFQCDdef}
\end{align}
where an average over color degrees of freedom is performed. Since the final state is necessarily attractive to support bound states and the gluon carries away an octet color charge, the initial scattering state is always in the repulsive adjoint representation. The same equation also holds for the more general $SU(N_c)$ gauge group, as well as for a complex scalar field (see \eg~Ref.~\cite{Binder:2021otw,Biondini:2023yxt} for the case of non-fundamental representations). For simplicity, when referring to the case $SU(N_c)$ in the following we mean either pairs of Fermions or complex scalars in the fundamental and anti-fundamental representation.

In the remainder of this section, we consider a constant coupling in the perturbative regime. We do this to investigate the unitarity violation independently of the effects induced by a running coupling. As long as the beta function is negative, running coupling effects can only enlarge the regime where partial wave unitarity is violated. We will return to the impact of running in the sections below.

\begin{figure}
    \centering
    \includegraphics[scale=0.73]{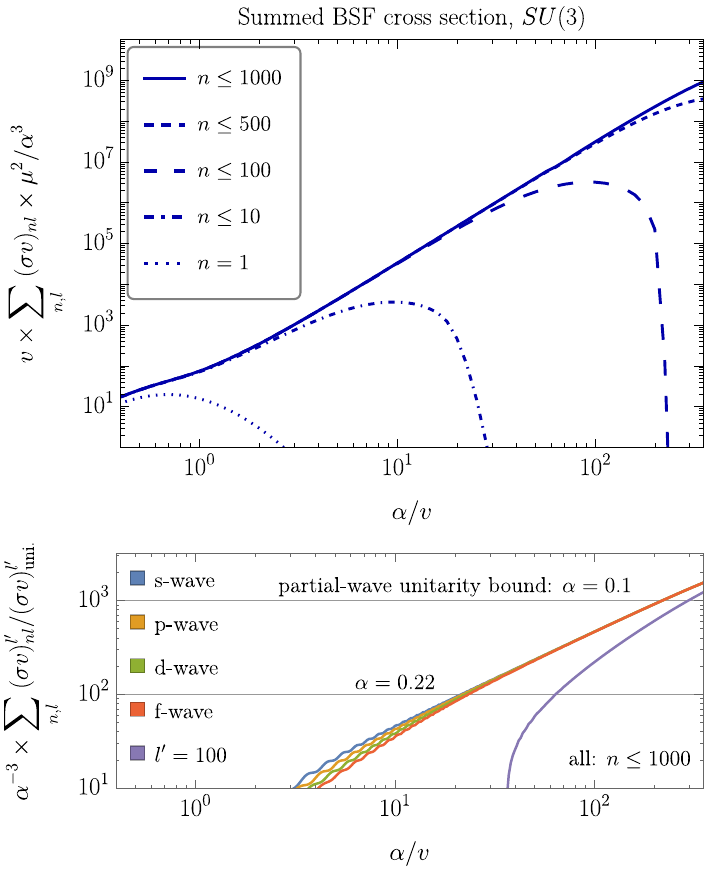}
    \caption{\emph{Upper:} The adjoint-to-singlet bound-state formation cross section in Eq.~(\ref{eq:sigmaBSFQCDdef}) shown for $SU(3)$, summed over all possible final bound-state quantum numbers $n$ and $\ell  \leq n-1$. From the truncation of the sum for $n\leq1,10,100,1000$ (blue), one can infer that higher excited states contribute more for decreasing velocity. \emph{Lower:} Rescaled ratio of BSF cross section for fixed initial angular momentum $\ell'$ and the corresponding partial-wave unitarity limit. For a given $\alpha$, unitarity is violated when the rescaled ratio is above $1/\alpha^3$, for which two examples are shown by the horizontal lines.}
    \label{fig:svQCD}
\end{figure}

Adopting this setting, we evaluate the $SU(3)$ BSF cross section in Eq.~(\ref{eq:sigmaBSFQCDdef}) for various $n,\ell$ and present the results in Fig.~\ref{fig:svQCD} in analogy to the previous case of $U(1)$. In the upper panel, one can notice that the velocity dependence of the summed BSF cross section may approach a power law when including as many excited states as our numerical limit allows for ($n\leq1000$, $\ell\leq n-1$) and considering only the region of $\alpha/v \lesssim 100$ where $n\leq1000$ are expected to be sufficient to capture the full result accurately. The scaling is much stronger than the previous Kramer's logarithm in $U(1)$. In fact, it is even stronger than $1/v^2$. Specifically, when fitting the summed cross section $\sum_{n,\ell} (\sigma v)_{n\ell} \propto v^{-\gamma}$ for $v\ll \alpha$ with a power law, we obtain $\gamma \approx 4.0$.

Such a velocity dependence raises concerns regarding the partial-wave unitarity. To investigate this issue, a scattering state with fixed \emph{initial} angular momentum, denoted by $\ell^\prime$, needs to be considered. From Eq.~(\ref{eq:sigmaBSFQCDdef}) we separate out this contribution by splitting Eq.~(\ref{eq:BSFMatrixelement}) into two contributions and denote the corresponding cross section by $(\sigma v)^{\ell^\prime}_{n\ell}$, where the superscript denotes the fixed initial scattering state angular momentum. For a given $\ell^\prime$, we then sum $(\sigma v)^{\ell^\prime}_{n\ell}$ over all possible final state quantum numbers $n,\ell$, compatible with the selection rules $\ell=\ell^\prime \pm 1$,

\be
(\sigma v)^{\ell^\prime} = \sum_{n,\ell}\,(\sigma v)^{\ell^\prime}_{n\ell}\,.
\ee
We require that this quantity respects the $\ell^\prime$th-partial wave unitarity bound. Specifically, as done in Ref.~\cite{vonHarling:2014kha,Smirnov:2019ngs,Baldes:2017gzw,Bottaro:2023wjv}, we consider the partial-wave unitarity cross section for $2\rightarrow2$ total inelastic collisions given by~\cite{Griest:1989wd}
\begin{align}
(\sigma v)^{\ell^\prime}_{\text{uni.}}= \frac{\pi(2\ell^\prime+1)}{\mu^2 v}.\label{eq:unitarityxs}
\end{align}
In the lower panel of Fig.~\ref{fig:svQCD}, we show the introduced summed $ (\sigma v)^{\ell^\prime}$ divided by the corresponding $\ell^\prime$th-partial wave unitarity cross section from Eq.~(\ref{eq:unitarityxs}). This ratio is multiplied by $\alpha^{-3}$, leaving only a dependence on $\alpha/v$. In this way, the partial-wave unitarity limits for a given $\alpha$ correspond to horizontal lines with values equal to $\alpha^{-3}$. We show $\ell^\prime=0$ (s-wave) to $\ell^\prime=3$ (f-wave)  BSF cross section summed in $n \leq 1000$ and one particular high initial angular momentum with $\ell^\prime=100$. The maximum principle quantum number is taken as our numerical limit ($n \leq 1000$), which is sufficient for the first four partial-wave summed cross sections to converge for $\alpha/v \lesssim 100$. From these results, one can infer that each summed $(\sigma v)^{\ell^\prime}$ grows faster than $1/v$, \ie~the unitarity limit will be exceeded at a finite velocity, respectively, for any $\ell^\prime$ we can resolve. This is independent of the chosen coupling strength. We point out that among all partial waves the s-wave unitarity bound is always violated at the largest velocities, though all curves approach a common behavior for decreasing velocity. We explicitly checked for the s-wave case that partial wave unitarity is even violated when including only a single, large $n$ bound state, implying that it is not the summation over all possible final states for a given $\ell^\prime$ which is problematic. 

Moreover, we observe the same situation for various $SU(N_c)$. To make an even more general statement, let us consider $(\sigma v)^{\ell^\prime}$ as a function of the ratio $\alpha_s^\text{eff}/\alpha_b^\text{eff}$. Now, $SU(N_c)$ is a special case which lies in the region $\alpha_s^\text{eff}/\alpha_b^\text{eff}=[- 1/3, 0 [$, where the lower limit corresponds to $N_c=2$ and the upper to the large $N_c$ limit. The $U(1)$ case corresponds to $\alpha_s^\text{eff}/\alpha_b^\text{eff}=1$. Our numerical results suggest that in the range  $\alpha_s^\text{eff}/\alpha_b^\text{eff} < 1$, $(\sigma v)^{\ell^\prime}$ scales stronger than $1/v$, while for $\alpha_s^\text{eff}/\alpha_b^\text{eff}\geq 1$ no evidence of partial wave-unitarity violation is found within our numerical boundaries.

Note that the mechanism behind the unitarization of BSF for $\alpha_s^\text{eff}/\alpha_b^\text{eff} < 1$ is still an open problem. While we point out this problem here for BSF via \emph{dipole transitions}, it is worth noting that a similar question has been recently raised for \emph{monopole transitions} in Ref.~\cite{Oncala:2019yvj} where partial-wave unitarity can be violated already for capture into the ground state level $n=1$.

For  non-numerical evidence of unitarity violation in non-Abelian gauge theories, a simple analytic expression would be warranted. We managed to get an approximate analytic result by taking two limits in Eq.~(\ref{eq:sigmaBSFQCDdef}): (i) $\alpha_s^\text{eff} \rightarrow 0$ and subsequently (ii) $\tilde\zeta_b \rightarrow \infty$ where $\tilde\zeta_b =\alpha_b^\text{eff}/(n v)$. Taking these limits, we obtain the result for the s-wave case in $SU(N_c)$ 
\begin{align}
(\sigma v)^{\ell^\prime=0}_{n,\ell=1} \simeq \frac{C_F}{N_c^2} \frac{4 \alpha}{3} \frac{32 \pi \alpha^\text{eff}_b}{\mu^2} n(n^2-1), \text{ for (i) and (ii).} \label{eq:analyticsun}
 \end{align}
The two limits are justified for relative velocities which fulfill the condition\footnote{For adjoint-to-singlet BSF in $SU(N_c)$ \mbox{$\alpha_s^\text{eff}=-\alpha/(2N_c)$} and $\alpha_b^\text{eff}=C_F\alpha$, where $C_F=(N_c^2-1)/(2N_c)$.}
\begin{align}
2\pi |\alpha_s^\text{eff}| \ll v \ll \frac{\alpha_b^\text{eff}}{2 n^2}.\label{eq:condition}
\end{align}
In this velocity regime, we compared our direct numerical evaluation of Eq.~(\ref{eq:sigmaBSFQCDdef}) to the analytical result in Eq.~(\ref{eq:analyticsun}) for a variety of $N_c$ and $n$ values and find very good agreement. The fact that the s-wave BSF cross section reaches a constant value for the above velocity regime is another non-trivial check of our numerical implementation also for very large $n$.

However, the velocity regime may be too restricted to analytically proof unitarity violation for contributions of a single $n$. Namely, while for $SU(N_c)$ the s-wave BSF cross section approaches the unitarity limit for increasing $n$, the velocity regime where the analytic expression is valid becomes smaller and eventually -- (very) close to the unitarity bound -- the condition in Eq.~(\ref{eq:condition}) cannot be met. Nevertheless, \emph{if} there exists a theory with $\alpha_s^\text{eff}=0$, then there is no lower bound on $v$ and violation of s-wave unitarity can be shown with the above formula. Notice that $\alpha_s^\text{eff}=0$ corresponds to the large $N_c$ limit of $SU(N_c)$, which is, however, not justified for all velocities for a finite $N_c$.

In the following, we explore phenomenological consequences focusing on the regime compatible with perturbativity and partial wave unitarity bounds.

%===================================================================
\section{Super critical behavior }\label{sec:darksectors}
%===================================================================

The impact of a set of bound states on the freeze-out dynamics of some particle species, $j$, can under very general conditions be described by the Boltzmann equation
\begin{align}
    \dot{n}_j + 3 H n_j= - \langle \sigma v \rangle_\text{eff} [ n^2_j-(n^\text{eq}_j)^2 ]\,,
    \label{eq:simpleBME}
\end{align}
where $n_j$ is the number density and $H$ the Hubble expansion rate. The \emph{effective cross section}, $\langle \sigma v \rangle_\text{eff}$, includes all the effects of pair annihilation as well as scattering-bound~\cite{Ellis:2015vaa} and bound-bound transitions~\cite{Mitridate:2017izz,Garny:2021qsr,Binder:2021vfo}. Here, we investigate whether the inclusion of an increasing number of excited states can lead to an effective cross section that grows sufficiently fast to maintain efficient depletion of the (comoving) particle number density and, hence, prevent the particle species  (\eg~dark matter) from freezing out. We call this condition a \emph{super critical} behavior.

To obtain the threshold for such a super critical behavior, let us  consider a typical scenario where a particle species with mass $m$ is initially in thermal equilibrium with a heat bath with temperature $T$ and entropy density $s$. We assume $s\propto T^3$, $H\propto T^2$, \ie~no (significant) change in the relativistic degrees of freedom of the bath. Introducing the yield as $Y_j\equiv n_j/s$ and parametrizing time by $x\equiv m/T$ in Eq.~\eqref{eq:simpleBME}, one can estimate the yield evolution as a function of $x$ as follows. For times where the yield $Y_j(x)$ starts to deviate significantly from its equilibrium value, $Y_j(x)\gg Y^\eq_j(x)$, also known as the time of chemical decoupling, $x_{\text{cd}}$%\drop{$\sim 25$)}
, one can neglect the impact of $Y^\eq_j(x)$ in the Boltzmann equation. This allows for an analytic solution for the yield evolution after chemical decoupling (see \eg~Ref.~\cite{Scherrer:1985zt}), which up to constants, can be estimated to scale as
\begin{equation}
Y_j(x_0)\propto \frac{1}{\int_{x_{\text{cd}}}^{x_0} \diff x \, x^{-2}\, \langle \sigma v \rangle_\text{eff} (x) }. 
\end{equation}
The integral converges for $x_0\to \infty$ only if $\langle \sigma v \rangle_\text{eff}  (x)$ grows slower than $x$ while for $\langle \sigma v \rangle_\text{eff} \propto x^\gamma$ with $\gamma\ge 1$
the integral diverges. Accordingly, the particle species only freezes out for  $\gamma< 1$ (typical WIMP) while the particle continues to deplete for $\gamma\ge 1$. The critical value $\gamma=1$ leads to logarithmic depletion and sets the threshold for what we define a super critical behavior.
Above this threshold, the evolution of the yield approaches the scaling $Y_j\propto x^{1-\gamma}$ for $x \gg x_\text{cd}$. In this case, the effective annihilation rate $\Gamma_\text{eff}\equiv n_j \langle \sigma v \rangle_\text{eff}$ is dynamically driven to be proportional to the Hubble rate $\Gamma_\text{eff}\propto H$.

In the presence of bound states, the effective cross section introduced above can be written as~\cite{Garny:2021qsr,Binder:2021vfo}
\be \label{eq:effgeneral}
  \avb{ \sigma v }_\text{eff} =   \avb{ \sigma v}_\text{ann} + \sum_{n,\ell} \avb{\sigma v}_{n\ell} R_{n\ell}\,,
\ee
where the first term is the usual pair annihilation cross section, thermally averaged. In all cases considered in this work, it includes the Sommerfeld effect~\cite{Hisano:2003ec, Hisano:2004ds}. The second term contains the thermal average of the BSF cross sections, denoted as $\avb{\sigma v}_{n\ell}$. The summation over all bound-state quantum numbers contains a dimensionless, temperature dependent quantity, which obeys $0 \leq R_{n\ell} \leq 1$.\footnote{Within the electric dipole approximation, bound states with different spin are not directly coupled to each other. We thus leave the spin sum implicit, see App.~\ref{sec:AppRates} for details.} Thus, the presence of bound states always increases the value of the effective cross section and could eventually lead to a super critical behavior. Introducing a simpler index to label a specific combination of quantum numbers, $i=(n\ell)$, $R_{i}$
can explicitly be written as~\cite{Garny:2021qsr,Binder:2021vfo}
\bea 
  R_i &\equiv& 1 - \sum_k (M^{-1})_{ik} \frac{ \Gamma_\text{ion}^k }{ \Gamma^k }\label{eq:Ri}\,,\\
  M_{ik} &\equiv& \delta_{ik} - \frac{ \Gamma_\text{trans}^{i\to k} }{ \Gamma^i }\label{eq:Mij}\,,\\
  \Gamma^i &\equiv& \Gamma_\text{ion}^i + \Gamma_\text{dec}^i + \sum_{k\neq i}\Gamma_\text{trans}^{i\to k}\,.\label{eq:Gammai}
\eea
The last line defines the total width of a particular bound state. It consists of the ionization rate, the rate of decay (via annihilation of the bound state's constituents), and bound-to-bound transition rates, respectively. The latter contains bound state excitation and de-excitation rates.
In practice, we use the Milne relation (\cf~App.~\ref{sec:ThAvMilne}) to obtain the excitation rate from the de-excitation rate, and $\Gamma_\text{ion}^{n\ell}$ from $\avb{\sigma v}_{n\ell}$. Note that the inclusion of bound-to-bound transition rates  \emph{increases} $\sum_{n,\ell}\avb{\sigma v}_{n\ell} R_{n\ell}$~\cite{Binder:2021vfo}.

%-------------------------------------------------------------------
\subsection{Dark QED}\label{sec:darkQED}
%-------------------------------------------------------------------

We now investigate the behavior of the effective cross section in Eq.~(\ref{eq:effgeneral}) for our first example of a concrete model. In particular, we consider dark matter as a Dirac Fermion charged under a $U(1)$ gauge group, which has been studied \eg~in Refs.~\cite{vonHarling:2014kha, Kamada:2019jch,Biondini:2023zcz}. We shall call it dark QED in the following, where dark photons set the thermal environment with temperature $T$.  

Dark QED has only two parameters, which are the dark matter mass, $m$, and the dark fine structure constant, $\alpha$. For our analysis, we consider Sommerfeld enhanced annihilation and s-wave spin-singlet bound state decay into two dark photons. Relevant expressions are listed in App.~\ref{sec:AppDarkQED}. We briefly comment on the influence of spin-triplet states below. The electric dipole interaction allows for transitions among the excited states in dark QED\@. The de-excitation rate is given by
\begin{align}\label{eq:GammaTrans}
    \Gamma_{\text{de-ex}}^{n^\prime \ell^\prime \rightarrow
    n \ell} = \frac{4 \alpha (2\ell +1)}{3} \Delta E^3 |\bra{\psi_{n\ell}} \mathbf{r} \ket{\psi_{n^\prime \ell^\prime}}|^2,
\end{align}
see App.~\ref{sec:btb}. The excitation rate is related via detailed balance, see App.~\ref{sec:AppRates}. 

\begin{figure}
    \centering
    \includegraphics[scale=0.74]{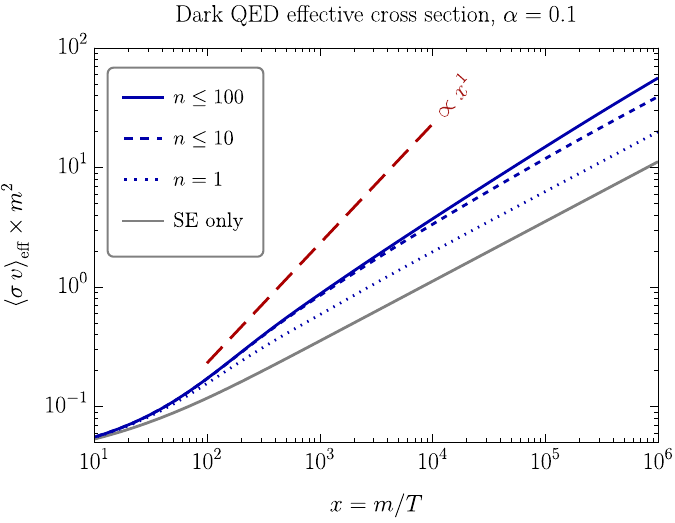}
    \caption{
    \label{fig:avQED}
    Effective cross section for heavy Fermions charged under a $U(1)$ for constant coupling $\alpha=0.1$, including the contribution from bound states for all $n,\,\ell$ up to $n=1,10,$ and $100$ (blue dotted, dashed, and solid, respectively) and excluding BSF (gray solid). The effective cross section includes Sommerfeld enhanced annihilation, BSF, ionization, and all possible bound-to-bound transitions arising from the electric dipole interaction, as well as spin-singlet s-wave bound state decay.
    The red long-dashed line displays the slope $\propto x^1$ for comparison.
    }
\end{figure}

Taking into account all leading processes, we show our results for the dark QED effective cross section in Fig.~\ref{fig:avQED}. Let us start by neglecting all bound state contributions, considering the case of Sommerfeld enhanced dark matter pair annihilation into two dark photons only (gray line). As is well known, the Sommerfeld effect in this case introduces a $1/v$ dependence of the annihilation cross section for $v \ll \alpha$, leading to $\langle \sigma v \rangle_\text{eff} \propto x^{1/2}$ for sufficiently low temperatures. Next, we add the contribution of the spin-singlet ground state (blue dotted). 
Similarly to the Sommerfeld effect, the cross section for capture into the ground state also scales as $1/v$ for $v \ll \alpha$ (as seen in Fig.~\ref{fig:svQED}). For $T$ much lower than the binding energy, this leads to $\avb{\sigma v}_{10} \propto  x^{1/2}$. In this regime, the spin-singlet decay rate is much faster than the ionization rate due to Boltzmann suppression and consequently $R_{10} \rightarrow 1$, resulting again in an overall $\langle \sigma v \rangle_\text{eff} \propto x^{1/2}$ scaling. Compared to the Sommerfeld enhanced pair annihilation only, the effective cross section is larger by a constant factor in the low temperature regime, as expected~\cite{vonHarling:2014kha}.

\begin{figure}
    \centering
    \includegraphics[scale=0.74]{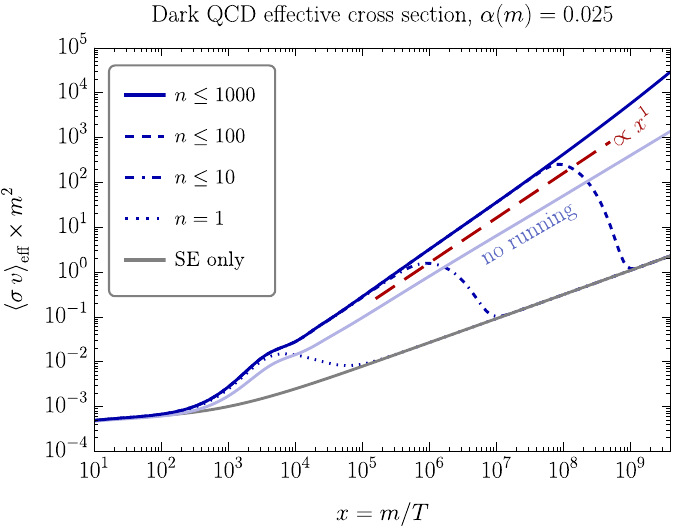}
    \caption{
    Effective cross section for heavy Fermion triplets under $SU(3)$, assuming $\alpha(m)=0.025$. The critical scaling $\langle \sigma v \rangle_{\text{eff}}\propto x^1$ (red dashed) is exceeded when respecting running couplings (darker blue), \ie~no freeze out occurs, as opposed to using constant coupling strength (lighter blue). 
    The effective cross section includes Sommerfeld enhanced annihilation, BSF and ionization via chromo electric dipole interactions, as well as spin-singlet s-wave bound state decay. Note that no bound-to-bound transitions occur in dark QCD in dipole approximation. The gray line shows the case without including bound states.
    }
    \label{fig:avQCD}
\end{figure}

Let us finally add many excited spin-singlet states to the system. We include, according to the selection rules, \emph{all} possible electric dipole transitions among them via Eq.~(\ref{eq:GammaTrans}) to evaluate the effective cross section in Eq.~(\ref{eq:effgeneral}). In Fig.~\ref{fig:avQED}, it can be seen that for $n \leq 10$ and $\ell\leq n-1$ (dashed blue line) the effective cross section increases strongly 
until around $x\sim 10^3$, although within this regime also even higher excited states become important as seen by the solid and dashed lines separating. One could therefore not deduce that dark QED does not exceed the critical power scaling $\gamma=1$ (indicated by the dashed red line) when including only $n\leq10$.
However, the result for $n \leq 100$ and $\ell\leq n-1$ (blue solid line), which includes about 5000 bound states and about $10^6$ transitions among them, clearly shows that the effective cross section does not continue a strongly increasing trend but rather converges to a smaller power scaling. When fitting a power law in the regime $10^4 \lesssim x \lesssim 10^5$, where $n \leq 100$ is trustworthy, we get a scaling of about $\langle \sigma v \rangle_\text{eff} \propto x^{0.6}$. Note that from the Kramer's logarithm in Sec.~\ref{sec:VelocityDependence}, it is not clear that the temperature dependence actually follows a power law.

The no-transition limit follows closely, but lies slightly above, the $n=1$ line as we explicitly checked. From this we conclude that transitions among the bound states are an important effect, which needs to be taken into account for predicting the relic abundance in dark QED precisely.

We verified, by varying the included $n$ between the two shown cases, that this less steep scaling, $\langle \sigma v \rangle_\text{eff} \propto x^{0.6}$, is already found for including only $n\leq 80$, 
from which we deduce that the inclusion of even higher excited states would not change the scaling. The scaling power is trivially unaffected by the value of the dark matter mass, as the mass cancels out in the shown product $\langle \sigma v \rangle_\text{eff} \times m^2$.
Moreover, one can show analytically that the scaling power is even unaffected when changing the value of $\alpha$, which we have confirmed numerically in a wide range of $\alpha$.
We also explicitly checked that the inclusion of spin-triplet bound states leaves the scaling power of the effective cross section at low temperatures unaffected.
This numerical observation can be understood since they only differ from the spin-singlet bound states by a smaller decay rate which suppresses their contribution at small $x$ but does not alter the late time scaling of the effective cross section (although it does increase the overall magnitude at late times).

From all this, we conclude that dark QED \emph{does not} reach a critical scaling of the effective cross section in the low temperature regime within the electric dipole approximation. In other words, dark QED indeed freezes out within the approximations made.\footnote{Interestingly, when considering processes with infinitely many dark photons (classical picture), other works come to a different conclusion~\cite{Belotsky:2015osa,Konstantin:2020rfx}. We only included ultra-soft processes with one dark photon.} We show the impact of excited states on the relic abundance within dark QED in App.~\ref{sec:uniQED}, refining earlier results on this subject~\cite{vonHarling:2014kha,Smirnov:2019ngs}.

%-------------------------------------------------------------------
\subsection{Dark QCD}\label{sec:darkQCD}
%-------------------------------------------------------------------

As our second example of a concrete model, we consider dark matter as a Dirac Fermion in the fundamental representation of a new $SU(3)$ gauge group, see \eg~\cite{Asadi:2021pwo,Biondini:2023zcz}, which is often called dark QCD. In the following analysis of dark QCD, we include standard expressions for the Sommerfeld enhanced pair annihilation cross section and the decay rate of the color singlet s-wave states, listed in App.~\ref{sec:AppDarkQCD} neglecting spin-triplet states. Further, for the octet-to-singlet BSF cross section in the chromo electric dipole approximation we use Eq.~\eqref{eq:sigmaBSFQCDdef}. No singlet-to-singlet transitions can be mediated via the chromo electric dipole operator. Therefore, the effective cross section in Eq.~(\ref{eq:effgeneral}) reduces to the no-transition limit $\Gamma_\text{trans}^{i\to j}=0$ (\cf~App.~\ref{sec:AppDarkQCD}) when allowing only for the leading chromo electric dipole interactions. This simplification allows us to 
focus exclusively on s-wave bound states which are the only ones with a non-vanishing decay rate in our approximations.

While in dark QED the coupling is frozen, in dark QCD, we take into account the one-loop running effect induced by gluon self-interactions in all considered quantities, as detailed in App.~\ref{sec:AppDarkQCD}.
Yet, $m$ remains the only dimensionful scale in the theory, implying that $\left< \sigma v\right>_\text{eff} \times m^2$ is independent of the choice of $m$. 

In Fig.~\ref{fig:avQCD}, we show  $\left< \sigma v\right>_\text{eff} \times m^2$  for the specific choice $\alpha(m)\equiv0.025$ across the regime $\alpha(m/x)\lesssim 1$. We checked that the BSF cross sections are compatible with partial wave unitarity bounds for the velocities that give a sizeable contribution to the thermal average, within the perturbative regime. The scaling power in the absence of bound states (Sommerfeld only) is, unsurprisingly, the same as in the dark QED case at low temperatures, \ie~$\left< \sigma v\right>_\text{eff} \propto x^{1/2}$. The inclusion of s-wave bound states, however, leads to a much steeper scaling of the effective cross section. Even when omitting the running of the coupling strength, \ie~$\alpha=0.025$ at all scales (light blue line, using $n\leq 1000$), we get a scaling power of about $x^{0.9}$ for $x > 10^5$, which is close to the critical line.
When fully including one-loop running (darker blue lines), the scaling of the effective cross section for dark QCD becomes super critical with a scaling of around $x^{1.1}$. 

We also find super critical behavior for other choices than $\alpha(m)\equiv0.025$. From this we conclude that dark QCD does not freeze-out even in the perturbative regime. However, since the scaling exceeds the critical one only slightly, we find a moderate effect on the abundance. For instance, assuming the dark sector is in thermal equilibrium with the SM bath, the addition of excitations $2\leq n\leq1000$ leads to a reduction of the dark matter abundance by around $50\%$ at $x=10^9$ for $m \sim 10^6\,{\rm{GeV}}$ and a slightly larger percentage for smaller masses.

In the following sections, we consider a model featuring dark matter and an accompanying particle that is charged under SM QED and QCD, and hence subject to bound state effects. The electric charge of that particle will allow for color singlet-to-singlet transitions, implying that the inclusion of $\ell >0$ states pushes the scaling of the effective cross section further inside the super critical regime~(\cf~Fig.~\ref{fig:svThAv}).  As a consequence, the corrections to the relic abundance in the perturbative regime will be much larger.

%===================================================================
\section{Colored t-channel mediator model}\label{sec:toymodel}
%===================================================================

We consider  a singlet Majorana Fermion $\chi$ being the dark matter candidate, and a  scalar mediator $\tilde q$ with gauge quantum numbers identical to those of either an up- or down-type right-handed  SM quark $q_R$ (we focus on the latter case in our numerical results for concreteness). The dark matter field $\chi$ interacts with the SM only via the Yukawa interaction
\be
  {\cal L}_{\text{int}} = \lambda_\chi  \tilde q  \bar q_R  \chi + \text{h.c.}\,,
\ee
while the mediator $\tilde q$ has additional interactions with the $SU(3)_c$ and $U(1)_Y$ SM gauge fields given by the usual kinetic term with covariant derivatives. We assume a mass $m_{\tilde q}>m_\chi$ such that the mediator can decay into the dark matter particle rendering only the latter stable on cosmological time scales. This model belongs to the class of so-called $t$-channel mediator models, see \eg~\cite{Garny:2015wea}, that are being actively considered in the context of LHC dark matter searches, see \eg~\cite{Arina:2023msd}  for a recent account on the subject.\footnote{The term \emph{mediator} refers to the dark matter-SM interaction, not to be confused with the long-range force carrier which in this case is the gluon (and photon). Occasionally, this class of models has alternatively been dubbed \emph{charged parent particle model}.}

Dark matter production is governed by the interaction of $\chi$ with the mediator field $\tilde q$, controlled by $\lambda_\chi$, as well as the dynamics of the mediator itself, largely driven by its 
gauge interactions. In particular, as the $\tilde q$ and $\tilde q^\dag$ particles are color and electrically charged, they can form bound states, that have an important impact on the freeze-out~\cite{Mitridate:2017izz,Biondini:2018pwp,Harz:2018csl,Biondini:2018ovz,Decant:2021mhj,Bollig:2021psb,Garny:2021qsr,Becker:2022iso}. Here, we are particularly interested in including excited bound states, following~\cite{Harz:2018csl,Garny:2021qsr}.

%-------------------------------------------------------------------
\subsection{Review of production mechanisms}\label{sec:production}
%-------------------------------------------------------------------

There are three distinct possibilities for  how the freeze-out dynamics occurs, known as coannihilation~\cite{Griest:1990kh,Edsjo:1997bg}, conversion-driven freeze-out~\cite{Garny:2017rxs,DAgnolo:2017dbv} and superWIMP production~\cite{Covi:1999ty,Feng:2003uy}, respectively. In addition, the model can also feature freeze-in production~\cite{McDonald:2001vt,Asaka:2005cn,Hall:2009bx}. 

All of them can be described by the following set of coupled Boltzmann equations for the yields $Y_j$, 
\begin{align}
  &\frac{ \diff Y_{\tilde q } }{\diff x}  = \frac{1}{3H}\frac{\diff s}{\diff x}
  \Bigg[ \frac{1}{2}\avb{ \sigma_{\tilde q\tilde q^\dagger}v }_\text{eff} \left(Y_{\tilde q}^2-Y_{\tilde q}^{\eq\,2}\right)\label{eq:BEqtilde}\\
  &\;+\avb{ \sigma_{\chi \tilde q}v } \left(Y_{\chi}Y_{\tilde q}-Y_{\chi}^{\eq}Y_{\tilde q}^{\eq}\right) 
  +\frac{\Gamma_\text{conv}^{\tilde q\to\chi}}{s} \left(Y_{\tilde q}-Y_{\chi}\frac{Y_{\tilde q}^{\eq}}{Y_{\chi}^{\eq}}\right)
  \Bigg]\,, \nonumber\\
  &\frac{\diff Y_{\chi }}{\diff x}  =   \frac{1}{ 3 H}\frac{\diff s}{\diff x}
  \Bigg[ \avb{ \sigma_{\chi\chi}v } \left(Y_{\chi}^2-Y_{\chi}^{\eq\,2}\right)\label{eq:BEchi}\\
  &\;
  +\avb{ \sigma_{\chi\tilde q} v } \left(Y_{\chi}Y_{\tilde q}-Y_{\chi}^{\eq}Y_{\tilde q}^{\eq}\right)
  -\frac{\Gamma_\text{conv}^{\tilde q\to\chi} }{s} \left(Y_{\tilde q}-Y_{\chi} \frac{Y_{\tilde q}^{\eq}}{Y_{\chi}^{\eq}}\right) 
  \Bigg] \nonumber\,,
\end{align}
where $x=m_{\tilde q}/T$, $s$ is the entropy density, $H$ the Hubble rate, and 
\begin{equation}
Y_j^\eq=\frac{g_j}{s}\int \frac{\diff^3p}{(2\pi)^3}\E^{-\sqrt{m_j^2+p^2}/T},
\end{equation}
all of which depend on $x$. Here $g_j$ denotes the
number of internal degrees of freedom, with $g_{\tilde q}\equiv 2N_c=6$ denoting the sum of $\tilde q$ and $\tilde q^\dag$ densities, $g_\chi=2$, and $g_{{\cal B}_{n\ell}}=2\ell+1$ for bound states with angular momentum $\ell$, capturing the degenerate magnetic quantum number. Note that the factor of $1/2$ in Eq.\,\eqref{eq:BEqtilde} is due to our convention of including both the $\tilde q$ and $\tilde q^\dag$ density in $Y_{\tilde q}$.

Eqs.\,\eqref{eq:BEqtilde} and \eqref{eq:BEchi} contain the following collision terms:
\begin{enumerate}
\item The effective cross section $\avb{ \sigma_{\tilde q\tilde q^\dagger}v }_\text{eff}$ includes direct annihilation (including Sommerfeld enhancement following~\cite{Ibarra:2015nca,Garny:2017rxs}) as well as the impact of bound states,
as given by Eq.~\eqref{eq:effgeneral}. We discuss the relevant BSF, transition and decay processes within the
simplified model below, following and extending~\cite{Garny:2021qsr}.
\item The rate $\Gamma_\text{conv}^{\tilde q\to\chi}$ describes the conversion rate of $\tilde q$ into $\chi$ particles. It is controlled by the Yukawa coupling, $\Gamma_\text{conv}^{\tilde q\to\chi}\propto \lambda_\chi^2$, and its size determines whether the freeze-out happens in the coannihilation, conversion-driven or superWIMP regime (see below). At high temperatures, it is dominated by scatterings $X\tilde q\to Y\chi$ with appropriate SM particles $X, Y$, while at low temperatures the decay process $\tilde q\to q \chi$ dominates. Accordingly, in the low temperature limit -- relevant for the superWIMP mechanism considered below -- it reads
\be
   \Gamma_\text{conv}^{\tilde q\to\chi} = \Gamma_{\tilde q\to q\chi}\frac{K_1\left( m_{\tilde q} /T\right) }{ K_2 \left( m_{\tilde q} /T \right) }\,,
\ee
where $\Gamma_{\tilde q\to q\chi}= \lambda_\chi^2/(16\pi)\, m_{\tilde q} \left(1-m_\chi^2/m_{\tilde q}^2\right)^2$ is the vacuum decay rate of a single mediator particle in the limit $m_q\to 0$ (not to be confused with the bound state decay rates, that are dominated by the strong interaction).
\item The dark matter pair annihilation rate $\avb{ \sigma_{\chi\chi}v }\propto \lambda_\chi^4 $ and coannihilation rate $\avb{ \sigma_{\chi\tilde q} v }\propto \lambda_\chi^4 $ are strongly suppressed for $\lambda_\chi\ll 1$ and practically irrelevant within the conversion-driven and superWIMP regimes.
\end{enumerate}
In addition, the Boltzmann equations could be complemented by collision terms for the conversion process $\tilde q\tilde q^\dag\to\chi\chi$, which are, however, negligible within the conversion-driven and superWIMP regimes as they are, again, proportional to $\lambda_\chi^4$, and irrelevant in the coannihilation limit, and therefore not displayed here.

\smallskip

Let us now discuss in more detail the various possible regimes for dark matter genesis. As mentioned above, which regime is realized depends on the size of the conversion rate.
More precisely, the most relevant quantity is the conversion rate for $\chi$ into $\tilde q$ particles, 
\be
  \Gamma_\text{conv}^{\chi\to\tilde q} = \Gamma_\text{conv}^{\tilde q\to\chi}\frac{Y_{\tilde q}^{\eq}}{Y_{\chi}^{\eq}} \to \Gamma_{\tilde q\to q\chi} \frac{g_{\tilde q}m_{\tilde q}^{3/2}}{g_\chi m_\chi^{3/2}}e^{-\frac{m_{\tilde q}-m_\chi}{T}}\,,
\ee
where the last expression is the low temperature limit. Dark matter genesis is qualitatively different depending on the size of this rate relative to the Hubble rate for temperatures around the mediator mass (during the time when the mediator starts to chemically decouple from the SM bath), 
\begin{subequations}
\label{eq:regimes}
    \begin{align}
  \Gamma_\text{conv}^{\chi\to\tilde q} \gg H (m_{\tilde q})&& \text{coannihilation}\,,\\
  \Gamma_\text{conv}^{\chi\to\tilde q} \sim H(m_{\tilde q}) && \text{conversion-driven}\,,\\
  \Gamma_\text{conv}^{\chi\to\tilde q} \ll H(m_{\tilde q}) && \text{superWIMP/freeze-in}\,.
    \end{align}
\end{subequations}
\begin{enumerate}
\item[(a)] In the coannihilation regime the ${\tilde q}$ and $\chi$ populations are in mutual chemical equilibrium. The actual size of the conversion rate is irrelevant as long as it is strong enough to maintain chemical equilibrium~\cite{Griest:1990kh,Edsjo:1997bg}. Within the coannihilation regime, the dark matter abundance is determined by the cross sections $\avb{\sigma_{\tilde q\tilde q^\dagger}v}$, $\avb{ \sigma_{\chi\chi}v }$ and $\avb{\sigma_{\chi\tilde q}v}$ (and in addition, as for all cases, the bound state dynamics), and generically here $\lambda_\chi\sim {\cal O}(1)$~\cite{Garny:2015wea,Harz:2018csl,Garny:2021qsr,Becker:2022iso}.

\item[(b)] In the conversion-driven case, the freeze-out of chemical equilibrium among $\chi$ and $\tilde q$ drives the dynamics and the size of the conversion rate $\Gamma_\text{conv}^{\tilde q\to\chi}$ largely influences the dark matter abundance~\cite{Garny:2017rxs,DAgnolo:2017dbv}. In addition, the efficiency by which the $\tilde q$ (and $\tilde q^\dag$) abundance is depleted is relevant~\cite{Garny:2017rxs}, controlled by $\avb{\sigma_{\tilde q\tilde q^\dagger}v}$ and bound state effects~\cite{Garny:2021qsr}. The conversion-driven case occurs for small couplings, typically $\lambda_\chi\sim {\cal O}(10^{-6})$, for which the $\chi\chi$ and $\chi\tilde q$ terms in Eqs.~\eqref{eq:BEqtilde} and \eqref{eq:BEchi} can be safely neglected.

\item[(c)] Finally, in the superWIMP scenario~\cite{Covi:1999ty,Feng:2003uy}, the mediator has an even smaller decay rate, and can usually be considered as stable while the freeze-out of $\tilde q\tilde q^\dag$ annihilation occurs. The population of remaining $\tilde q$ and $\tilde q^\dag$ particles then decays into $\chi$ at a temperature $T$ for which $H(T)\sim \Gamma_\text{conv}^{\chi\to\tilde q}$, thereby generating the dark matter abundance~\cite{Garny:2018ali}. Technically, this means that the terms corresponding to inverse decays in Eqs.~\eqref{eq:BEqtilde} and \eqref{eq:BEchi} can be neglected, in addition to those for $\chi\chi$ and $\chi\tilde q$ annihilation, while the size of $\avb{\sigma_{\tilde q\tilde q^\dagger}v}$ and the bound state dynamics are most important~\cite{Garny:2018ali,Decant:2021mhj,Bollig:2021psb}. To the extent that decay and freeze-out occur on different time-scales, the dark matter abundance is also insensitive to the size of the conversion (or equivalently decay) rate in that limit, since eventually each $\tilde q$ (and $\tilde q^\dag$) produces one dark matter particle. Note that in addition to the superWIMP contribution a contribution from freeze-in~\cite{McDonald:2001vt,Asaka:2005cn,Hall:2009bx} has to be considered which stems from inefficient decays (or scatterings) around $x\sim1$, \ie~when the mediator is in thermal equilibrium with the SM bath. The relative importance of superWIMP versus freeze-in contributions depend on the couplings and masses, see \eg~\cite{Garny:2018ali,Decant:2021mhj}. However, in our analysis we are particularly interested in regions with a dominant superWIMP contribution.
\end{enumerate}
In this work, we re-evaluate the superWIMP regime when including excited bound state effects. In particular, since the mediator is relatively long lived within this regime, its abundance crucially depends on how much of the mediator is depleted due to bound state dynamics.

%-------------------------------------------------------------------
\subsection{Bound state rates and processes}\label{sec:bound}
%-------------------------------------------------------------------

The impact of bound states on $Y_{\tilde q}$ is captured by the effective cross section defined in Eq.~\eqref{eq:effgeneral}  entering in the Boltzmann equation, Eq.~\eqref{eq:BEqtilde}. It depends on the set of bound states that are included as well as their formation, transition and decay rates, discussed in the following.

In the considered model, the scalar mediator particle $\tilde q$ interacts both via electromagnetic interaction with bottom-like charge $Q=-1/3$ and strong interactions in the fundamental representation. This leads to differences compared to the case of pure Abelian or non-Abelian interactions discussed in Sec.\,\ref{sec:darksectors}. In particular, the potentials determining the bound state spectrum and wave-functions as well as BSF and decay are driven by QCD, while QED is relevant for transitions among the various energy levels~\cite{Garny:2021qsr}. In the following, we briefly review the bound state processes included in our analysis, and then comment on the relevance of further extensions. 

Bound state formation is dominated by the chromo electric dipole transition,
\be
  (\tilde q\tilde q^\dag)^{[8]} \to {\cal B}_{n\ell}^{[1]} + g\,,
\ee
going from an octet scattering state to a singlet bound state, and emitting an (ultrasoft) gluon. The effective interaction potentials $V_{s(b)}=-\alpha_{s(b)}^\text{eff}/r$ for the scattering ($s$) and bound states ($b$) are
\be
  \alpha_{s}^\text{eff} = -\frac16\alpha_s(\mu_s),\quad
  \alpha_{b}^\text{eff} = \frac43\alpha_s(\mu_b)\,,
\ee
with running strong coupling\footnote{In all numerical computations involving running $\alpha_s$ we used \textsc{RunDec}~3~\cite{Herren:2017osy} to evaluate the SM strong coupling (employing 5-loop running).} evaluated
at $\overline{\text{MS}}$-scale $\mu_s=m_{\tilde q} v_\text{rel}/2$ for the scattering state, and at the Bohr
momentum scale $\mu_b=m_{\tilde q} \alpha_{b}^\text{eff}/2/n$ for the bound state
${\cal B}_{n\ell}$. Note that the latter definition is implicit, but can be easily solved either iteratively or numerically for each level $n$. The BSF cross section is then given by Eq.~\eqref{eq:sigmaBSFQCDdef}, with the effective running coupling entering the respective initial and final state wave-functions. Furthermore, we evaluate the coupling in the prefactor of Eq.~\eqref{eq:sigmaBSFdef}, associated to the gluon emission, at the ultrasoft scale of the gluon energy $\mu_\text{BSF}=m_{\tilde{q}}/{4}\,\left(v^2+(\alpha_b^\text{eff}/n)^2\right)$. Bound state formation is also possible via an electromagnetic dipole transition, but is negligible due to the smaller interaction strength~\cite{Garny:2021qsr} (see also Fig.~\ref{fig:Triangle} below). Computational details can be found in App.~\ref{sec:AppDarkQCD}. 

Transitions among bound states are not possible via a single insertion of the QCD dipole interaction due to color conservation. Therefore, we consider electromagnetic dipole interactions as the leading bound-to-bound transitions:
\be\label{eq:EM transition}
  {\cal B}_{n\ell}^{[1]} \leftrightarrow {\cal B}_{n'\ell\pm1}^{[1]} + \gamma\,,
\ee
with rates computed as detailed in App.~\ref{sec:AppSWIMP}. We use the wave-functions evaluated with the corresponding effective {QCD} couplings at their respective Bohr momentum scales for level $n$ and $n'$ and the electromagnetic fine structure constant $\alpha_\text{EM}=1/128.9$ in the coupling prefactor of Eq.~\eqref{eq:GammaTrans}. 

For bound state decay we include the process \mbox{${\cal B}_{n,\ell=0} \to gg$}, see App.~\ref{sec:AppDarkQCD}. The next-to-leading order correction to this decay channel within QCD has been shown
to have only a minor effect in~\cite{Garny:2021qsr}, and we therefore
omit it here.

Let us briefly comment on possible further transition and decay processes.  By simple power counting, electric quadrupole and magnetic dipole transitions are suppressed relative to electric dipole transitions. Nevertheless, they could potentially have an impact by allowing new transition channels due to the modified selection rules. We checked (up to $n\leq 6$) that electric quadrupole transitions have a negligible impact on the effective cross section. Furthermore, the decay of $\ell=1$ bound states is suppressed by at least a factor $\alpha_s^3$ relative to its de-excitation rate when assuming a power counting $\alpha_\text{EM}\sim\mathcal{O}(\alpha_s^2)$. This is due to the higher derivatives of the radial wavefunction entering for higher $\ell$, as well as the fact that decays into two gluons vanish for $\ell=1$ states at tree-level, leading to additional suppression due to a three-gluon decay or two-gluon decay at one-loop~\cite{Cacciari:2015ela}. Lastly, two-gluon transitions in $SU(N_c)$ are expected to be suppressed by phase space factors and the repulsive potential of the necessarily adjoint intermediate state. Nevertheless, it would be interesting to investigate their impact in future work.

\begin{figure*}
  \centering
  \includegraphics[width=0.99\textwidth]{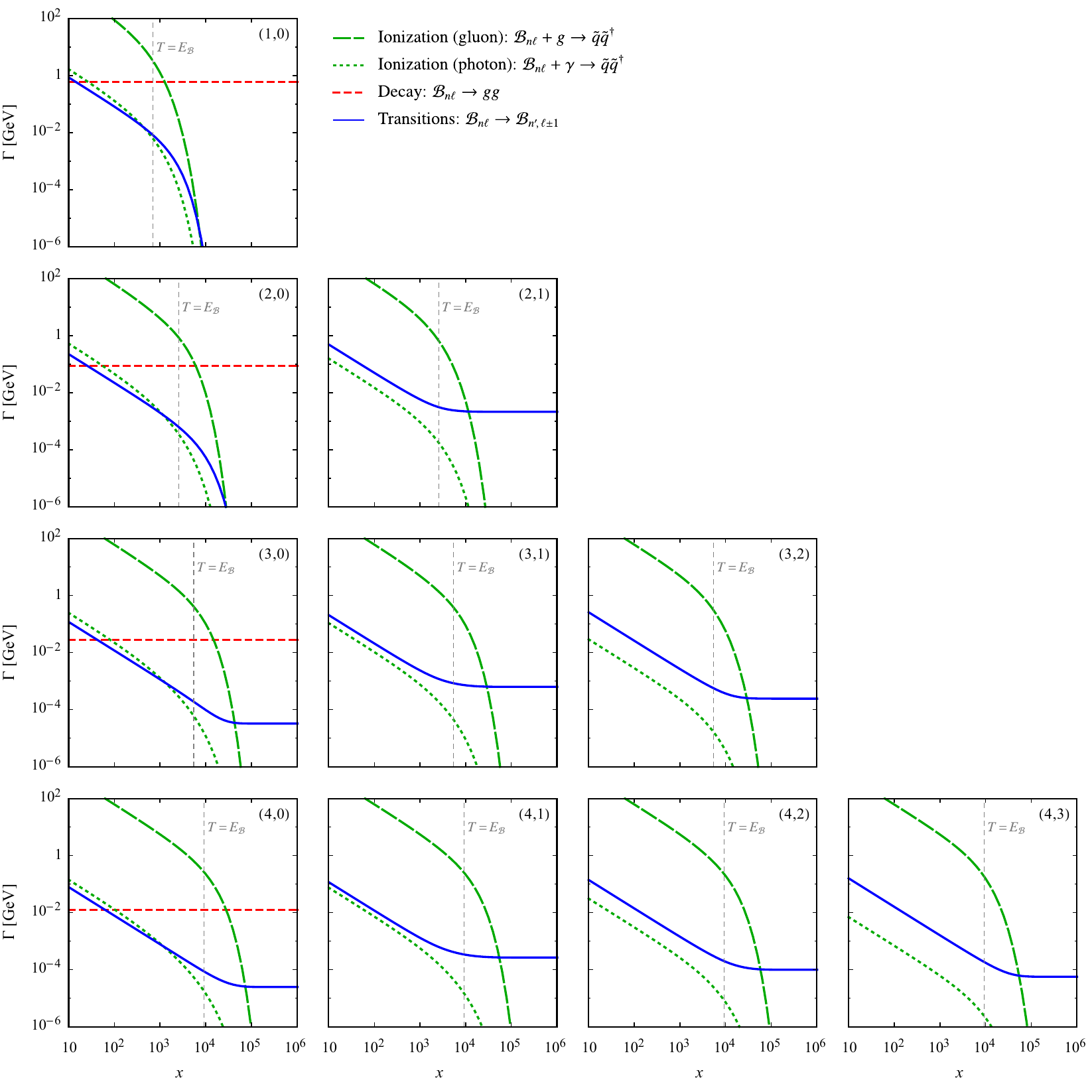}
  \caption{Ionization, decay and transition rates for all bound state levels $(n,\ell)$ up to $n=4$ for $m_{\tilde q}=4\times 10^6\,$GeV. The gray dashed vertical line indicates the temperature that corresponds to the binding energy $E_\mathcal{B}$ of the respective bound state. Note that the transition rates contain all possible excitations and de-excitations (here summed up to $n'=20$).
}
\label{fig:Triangle}
\end{figure*}

\begin{figure} 
    \centering
    \includegraphics[width=.48\textwidth]{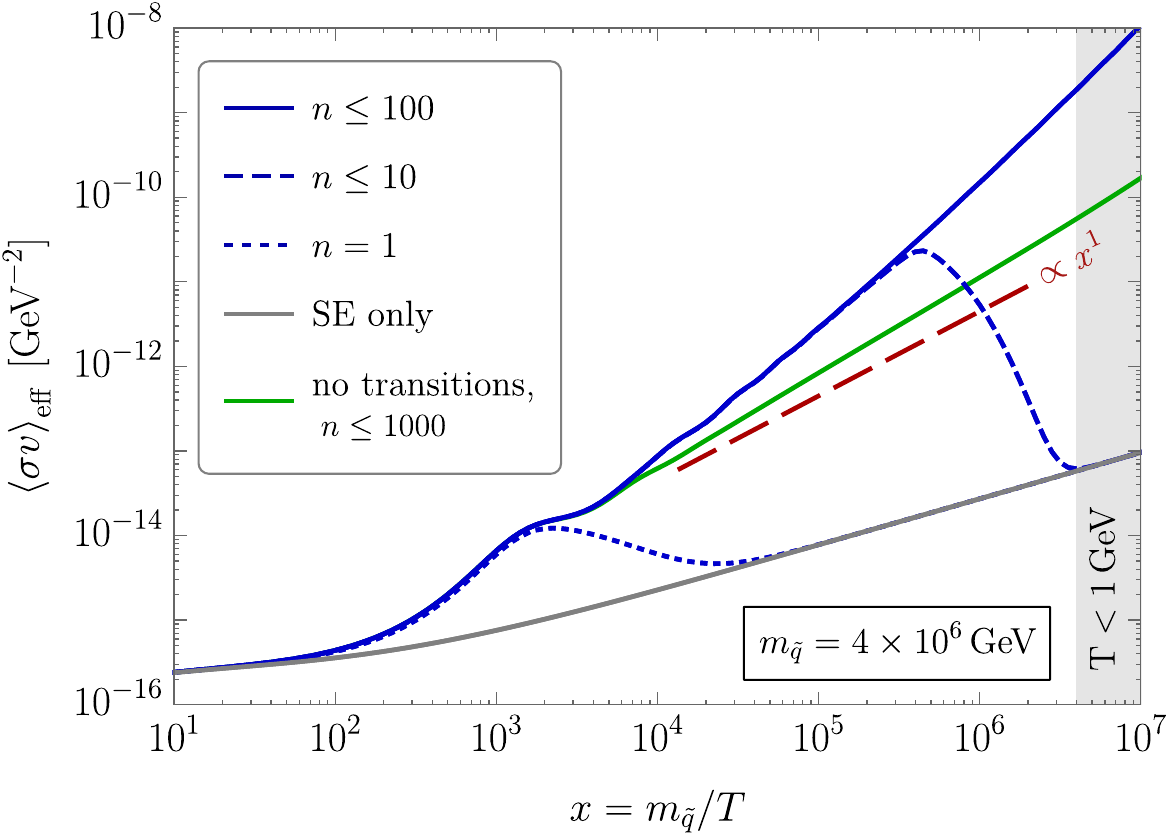}
    \caption{
    Effective cross section for the colored and electrically charged mediator $\tilde q$. It includes the contribution from bound state levels $(n,\ell)$ for all $\ell$ and up to $n\leq 1,10,100$ (blue lines) accounting for all dipole transitions among them. Also shown is the no-transition limit for $n\leq 1000$ (green). Both grow more steeply than  $\propto  x$ (indicated by a thin black line) when sufficiently high $n$ are included, implying mediator depletion without freeze-out. Here $m_{\tilde{q}}=4\times 10^6$GeV.}
    \label{fig:svThAv}
\end{figure}

%===================================================================
\section{Results for superWIMP scenario}\label{sec:swimp}
%===================================================================

We now discuss our results and phenomenological implications considering superWIMP production within the model introduced in Sec.~\ref{sec:toymodel}. 
In the superWIMP scenario, the standard assumption has been that the freeze-out and decay of $\tilde q$ are well separated in time, such that the late time $\chi$ abundance depends on the mediator freeze-out abundance only. The resulting dark matter density has thus been considered to be \emph{independent} of the mediator lifetime $\tau_{\tilde q}=1/\Gamma_{\tilde q\to q\chi}$. In our model, we will show that the superWIMP mechanism proceeds in a \emph{qualitatively different} way as the consequence of the super critical behavior of the effective cross section in non-Abelian gauge theories found in Sec.~\ref{sec:darkQCD}.

We assume that the mediator decays well before the QCD phase transition, such that the entire freeze-out dynamics takes place within the unconfined phase and involves $\alpha_s$ in the perturbative regime ($T>1$GeV) only. Numerical results presented throughout this section are shown for the representative benchmark point $m_{\tilde{q}}=4\times 10^6$~GeV. We     discuss the mass dependence of our findings in Sec.~\ref{sec:complementary}.

%-------------------------------------------------------------------
\subsection{Effective cross section}\label{sec:sigefftoy}
%-------------------------------------------------------------------

The abundance of the mediator $\tilde q$ is governed
by the effective cross section Eq.~\eqref{eq:effgeneral}, that encapsulates the impact of bound states. In Fig.~\ref{fig:Triangle}, we show the rates that enter this quantity for an exemplary subset with $n\leq 4$ and for all $\ell\leq n-1$, for a wide range of $x=m_{\tilde q}/T$ corresponding to $m_{\tilde q}/10\geq T\geq 4\,$GeV. Since the mediator is charged under both $SU(3)_c$ and $U(1)_Y$ it combines features of the Abelian and non-Abelian cases discussed in Sec.\,\ref{sec:darksectors}. 

As expected, ionization (and correspondingly BSF) is cleary dominated by the QCD-mediated process, as can be seen by comparing the long-dashed and dotted lines in Fig.~\ref{fig:Triangle}. Since $\tilde q\tilde q^\dag$ bound states exist only for the attractive color singlet configuration, color conservation dictates that bound-to-bound transitions are only contributing via an electric dipole interaction mediated by QED\@. The total transition rate from a given bound state $(n,\ell)$ into any higher or lower state is shown by the blue lines in Fig.~\ref{fig:Triangle}. For the ground state $(1,0)$ only excitation occurs such that the rate becomes exponentially Boltzmann suppressed once the temperature drops below the corresponding difference of binding energies. The same is true for $(2,0)$ due to the selection rule $\Delta\ell=\pm 1$ for dipole transitions. For all other levels, the total transition rate approaches a finite value for low temperature (\ie~large $x$), corresponding to the rate of de-excitation into lower levels. In addition, we include the direct decay of $\ell=0$ states into a pair of gluons, which is the analog of mediator pair annihilation. This rate is practically temperature independent for $x\gg 1$. Overall, the total width of any given level is dominated by the QCD-mediated ionization rate at temperatures $T$ above or around the binding energy. For much lower temperatures, decay dominates for $\ell=0$, and QED-mediated transitions (to lower levels) for $\ell\geq 1$.

The resulting effective cross section, which combines QCD mediated BSF and QED transitions among bound states, is shown in Fig.~\ref{fig:svThAv}, where we take excited states ${\cal B}_{n\ell}$ with $0\leq \ell\leq n-1$ and up to a given maximal $n$ into account. For the various blue lines, we include all possible electric dipole transitions. For any given temperature, the effective cross section converges when including a sufficient number of excited states. 
The reason for this convergence is that each excited state contributes in a limited temperature range only. While the underlying velocity dependence of the BSF cross section becomes increasingly complex at large $n$, a given bound state $n,\,\ell$  only starts to contribute significantly once the temperature drops down to roughly its respective binding energy, $T\sim E_{{\cal B}_{n\ell}}$.  In fact, this is important for the validity of the dipole approximation, as it ensures that the temperature is well below the typical momenta of bound states that contribute significantly \ie~below their respective Bohr momenta. For $T\ll E_{{\cal B}_{n\ell}}$ in contrast, its contribution is negligible due to the repulsive potential in the scattering state~\cite{Harz:2018csl, Garny:2021qsr}.

Introducing $x_n\equiv m_{\tilde q}/E_{{\cal B}_{n\ell}}$, we find \mbox{$x_1\simeq 7\times 10^2$}, \mbox{$x_{10}\simeq 5\times 10^4$}, \mbox{$x_{100}\simeq 3\times 10^6$} for our benchmark \mbox{$m_{\tilde q}=4\times 10^6$\,GeV}, which correspond to the $x$ values at which the respective excited levels are expected to start contributing significantly to the effective cross section. Overall, the lower the temperature (\ie~the higher the $x$), the more relevant higher excited levels become for achieving a converged effective cross section. We include states up to $n=100$, taking all transitions among them into account (blue solid line). We checked that this suffices to reach converged results within the perturbative regime. In particular, the difference between $n\leq 50$ and $n\leq 100$ is less than $0.2\%$ for $T>1\,$GeV.

As visible in Fig.~\ref{fig:svThAv}, the effective cross section (blue solid line) clearly shows a super critical behavior, \ie~the power scaling is significantly larger than $\propto x$ (red dashed). We stress that the interplay of bound states formed by the non-Abelian QCD interaction with transitions mediated by QED leads to a significant enhancement of the effective cross section compared to the limit of inefficient transitions, see green line in Fig.~\ref{fig:svThAv}. The latter shows the result when omitting transition processes. It is similar to the case of dark QCD discussed in Sec.~\ref{sec:darkQCD}\@. This shows that excited states play an even more prominent role for a mediator charged under both QCD and QED considered here. Nevertheless, the effective cross section increases more steeply than $\propto x$ even in the no-transition approximation due to running. All the same, when including transitions but \emph{neglecting} running, the slope is still steeper than $\propto x$, \ie~the presence of bound-to-bound transitions causes a super-critical behavior in our model even without running coupling effects.

%-------------------------------------------------------------------
\subsection{Relic abundance}\label{sec:relic}
%-------------------------------------------------------------------

\begin{figure}
  %\centering
  \includegraphics[width=0.45\textwidth]{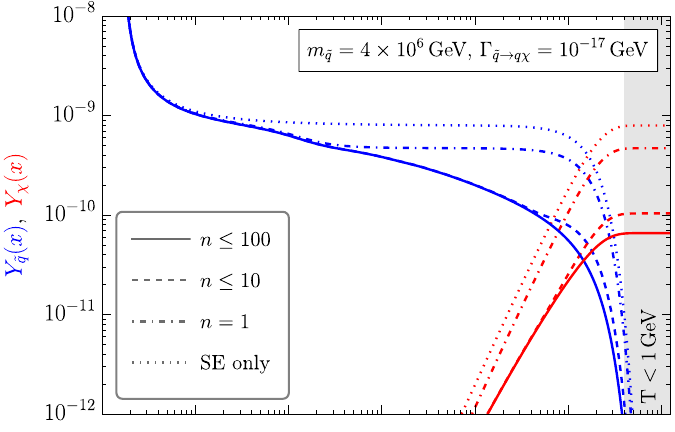} \\ [-4pt]
  \hspace{5pt}\includegraphics[width=0.442\textwidth]{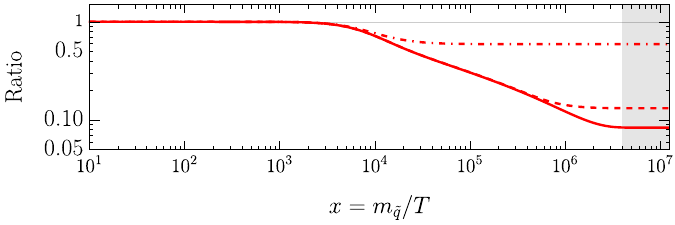}
  \caption{
 \emph{Upper panel:} Abundance evolution of $\chi$ (red) and $\tilde{q}$ (blue) for $m_{\tilde q}=4\times 10^6$\,GeV and $\Gamma_{\tilde q\to q\chi}= 10^{-17}$\,GeV. When including no (SE only) or one ($n=1$) bound state, the mediator yield $Y_{\tilde q}$ freezes out and then subsequently transfers its abundance to $Y_\chi$ via $\tilde q\to q\chi$. Taking excited states and transitions among them into account leads to a continuous decrease of $Y_{\tilde q}$ (dashed and solid blue line for $n\leq 10$ and $n\leq 100$, respectively) that is only terminated by mediator decay once $\Gamma_{\tilde q\to q\chi}\gtrsim H$. 
 \emph{Lower panel:} Ratio of  $\chi$ yield when including bound states up to $n\leq 1, 10, 100$ over the result without bound state contributions.
  }
  \label{fig:AbundanceEvolution}
\end{figure}
\begin{figure}
\centering
 \includegraphics[width=0.45\textwidth]{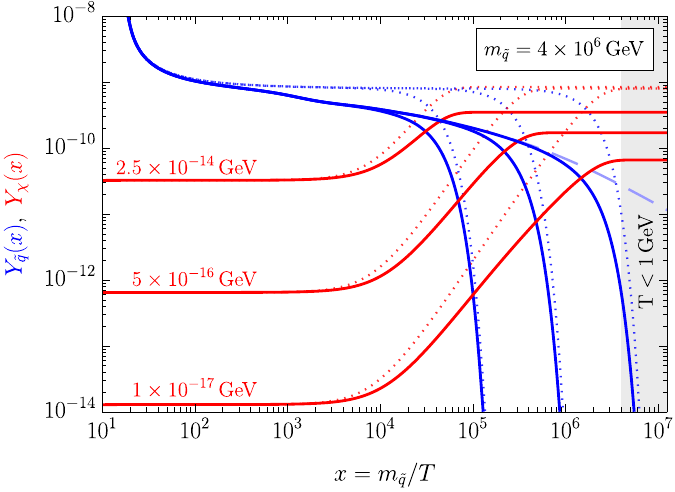}
\caption{
 Evolution of mediator and dark matter abundances when including bound states and transitions among them up to $n=100$ (solid lines) or no bound states (dotted) for three different decay rates $\Gamma_{\tilde q\to q\chi}=2.5\times 10^{-14}$, $5\times 10^{-16}$, $10^{-17}$\,GeV. The long dashed  line shows the limit $\Gamma_{\tilde{q}\to q\chi}=0$, which decreases due to the large contribution of excited states to the effective cross section. The final value of $Y_\chi$ therefore depends on $\Gamma_{\tilde q\to q\chi}$ even in the superWIMP regime where the $\chi$ abundance generated via freeze-in at low $x$ is negligible.
 }
 \label{fig:AbundanceMultiGamma}
\end{figure}

The evolution of the yields $Y_{\tilde q}(x)$ and $Y_\chi(x)$ for the mediator and the dark matter particle as obtained from solving the coupled Boltzmann equations \eqref{eq:BEqtilde} and \eqref{eq:BEchi} are shown in Fig.~\ref{fig:AbundanceEvolution}. Let us first discuss the case when including either only mediator annihilation (blue dotted line) or in addition the ground state (blue dot-dashed line, see \eg~\cite{Decant:2021mhj,Bollig:2021psb}). In these cases, the yield $Y_{\tilde q}(x)$ freezes out at $x\sim {\cal O}(10^2\!-\!10^3)$. After freeze-out it remains constant until the age of the Universe becomes comparable to the mediator lifetime, \ie~for $H\sim \Gamma_{\tilde q\to q\chi}$. At this point, the mediator particles decay into dark matter, such that the final yield $Y_\chi$ is identical to the freeze-out value of $Y_{\tilde q}$ which is set at much earlier times already. Accordingly, in a wide range of lifetimes the dark matter abundance does not depend on the dark matter coupling. This qualitative picture of the superWIMP mechanism has widely been adopted  throughout the literature in the past.

Intriguingly, when including excited bound states, the super critical effective cross section leads to a continuous depletion such that $Y_{\tilde q}$ never freezes out. This can be seen in the solid line in Fig.\,\ref{fig:AbundanceEvolution}, where bound states up to $n\leq100$ are included. 
The depletion of the number density is dominated by the effective cross section, until the decay of the mediator, $\tilde q\to q\chi$, becomes efficient (here $x \gtrsim 10^6$). In fact, the  effective annihilation rate $\Gamma_{\text{eff}}= n_{\tilde q}\langle\sigma v\rangle_{\text{eff}}$ is kept on the edge of being efficient, \ie~$\Gamma_{\text{eff}}\sim H$, over the entire period of bound  state induced depletion. The dynamics is qualitatively different to the standard  picture as there is no temperature regime where the yield of the mediator has frozen out. 

The quantitative impact of bound states up to \mbox{$n\leq1,10,100$} on the dark matter abundance is explicitly shown in lower panel normalized by the result including only Sommerfeld enhancement and no bound states. Taking into account the ground state only yields a reduction by a factor $\sim 2$ in the final abundance. When considering the full bound state effects ($n\le100$), we find that the dark matter relic abundance is lowered by more than an order of magnitude. 

Note that the mediator decays before the QCD transition such that dark matter production takes place in the unconfined phase involving $\alpha_s$ in the perturbative regime. We explicitly verified that our results are insensitive to the behavior of the strong coupling at scales below $1\,$GeV. To check this we implemented different numerical prescriptions for treating the strong coupling at these low scales, and find that the final abundance is highly insensitive as long as the mediator lifetime ensures a decay before the QCD transition. Furthermore, we checked that 
our results are not influenced by contributions for which partial-wave unitarity is questionable,
as the corresponding velocities are not relevant for the thermally averaged effective cross section in the regime $T>1$\,GeV.

In Fig.\,\ref{fig:AbundanceMultiGamma}, we show the abundance evolution for three different values of the mediator lifetime. (The additional long-dashed line depicts the result when excluding mediator decays, \ie~$\Gamma_{\tilde{q}\to q\chi}=0$.) For all decay rates, superWIMP production provides the dominant contribution to the dark matter density. While for the largest decay rate, $\Gamma_{\tilde q\to q\chi}=2.5\times 10^{-14}$\,GeV, freeze-in still contributes almost 10\%, it is fully negligible for the smaller rates chosen. Due to the continuous depletion of $Y_{\tilde q}$ in the presence of excited states, the time of decay does, indeed, have an impact on the final dark matter abundance, as can be seen from the three different values of $Y_\chi$ (red lines) for $x\to \infty$. In contrast, when neglecting excited states (dotted lines), the final yield is identical for all three mediator decay rates. A similar behavior can be found when including only  a small number of bound states.

%-------------------------------------------------------------------
\subsection{Implications}\label{sec:complementary}
%-------------------------------------------------------------------

\begin{figure}
  \centering
  \includegraphics[width=0.48\textwidth,trim={0 0 0.28cm 0},clip]{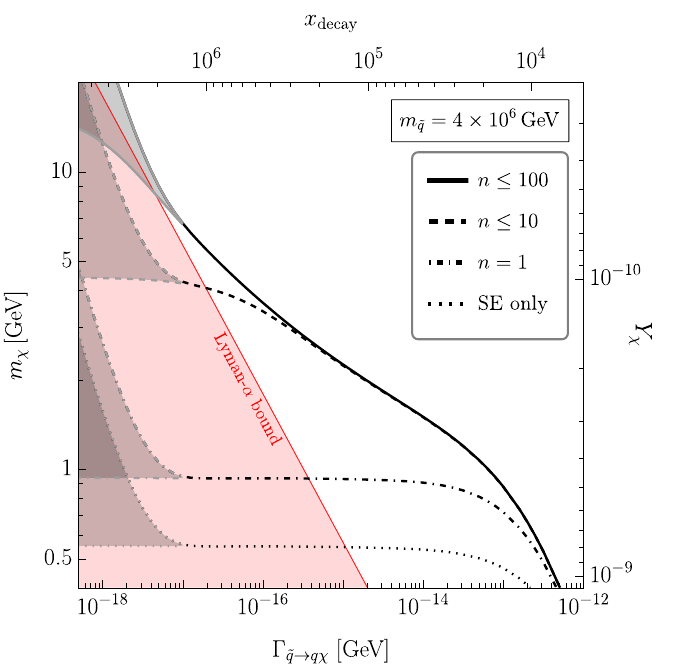}
  
  \caption{
  Dark matter mass $m_\chi$ and decay rate $\Gamma_{\tilde q\to q\chi}$ of the colored $t$-channel mediator for which the final yield matches the observed dark matter density $\Omega_{\chi}h^2=0.12$, using $m_{\tilde q}=4\times 10^6$\,GeV. The abundance is set by the superWIMP mechanism for $\Gamma_{\tilde q\to q\chi}\lesssim 10^{-13}$\,GeV, while freeze-in dominates for larger decay rates. When taking only Sommerfeld-enhanced $\tilde q\tilde q^\dag$ annihilation into account (dotted line) the relic density is independent of $\Gamma_{\tilde q\to q\chi}$ within the superWIMP regime, \ie~the dotted line is horizontal. The same is true when taking the ground state $n=1$ into account (dot-dashed line). When including excited states, the mediator depletion due to   BSF, transitions and decay of bound states
  continues  until eventually  $H\sim \Gamma_{\tilde q\to q\chi}$, such that the relic density does depend on the mediator lifetime, corresponding to the black dashed ($n\leq 10$) and black solid ($n\leq 100$) lines. For $\Gamma_{\tilde q\to q\chi}\gtrsim 10^{-17}$\,GeV the mediator decays before the QCD transition, \ie~within the perturbative regime. The shaded areas bracket the possible values for even lower decay rates, see text for details.
  The red shaded region is excluded at 95\%~C.L. by Lyman-$\alpha$ forest observations~\cite{Decant:2021mhj}.   
}
  \label{fig:eP}
\end{figure}

In Fig.~\ref{fig:eP}, we finally show the dark matter mass, $m_\chi$ (left axis labeling), for which the final $\chi$ abundance (displayed  using the right axis labeling) yields the observed dark matter relic density, $\Omega_\chi h^2\simeq 0.12$~\cite{Planck:2018vyg}, as a function of the mediator decay rate, $\Gamma_{\tilde q\to q\chi}$. For $\Gamma_{\tilde q\to q\chi}\gtrsim 10^{-12}$\,GeV, dark matter production is dominated by freeze-in,  while for lower decay rates, which we focus on in the following, the superWIMP mechanism sets the abundance. As discussed above, within this regime, the mediator abundance has so far usually been assumed to  freeze out. In that case, each remaining mediator particle subsequently produces one dark matter particle, such that the precise decay rate is irrelevant for superWIMP production. This is indeed the case in Fig.\,\ref{fig:eP}, when taking only mediator pair annihilation (dotted) or in addition the ground state (dot-dashed) into account.

When including excited bound states, the mediator abundance continues to deplete
until the age of the Universe reaches the mediator lifetime. The remaining mediators then produce dark matter via $\tilde q\to q\chi$. This implies that the final dark matter abundance does, in fact, depend on $\Gamma_{\tilde q\to q\chi}$ and, hence, on the dark matter coupling. Therefore also the dark matter mass $m_\chi$ for which $\Omega_\chi h^2\simeq 0.12$ does depend on it. This can be seen in the solid and dashed curves in Fig.\,\ref{fig:eP}  corresponding to taking into account all excited states ($\ell\leq n-1$) with $n\leq 10$ and $100$, respectively. The smaller dark matter abundance implies a larger dark matter mass, as compared to the case without excited states. 

For $\Gamma_{\tilde q\to q\chi}\gtrsim 10^{-17}$\,GeV, the mediator decay occurs prior to the QCD transition, \ie~within the perturbative regime $T>1$\,GeV\@.
Note that in this regime the inclusion of bound state up to $n=100$ appears to be sufficient. While we find significant contributions from bound states with $n>10$ (\cf~the difference between the dashed and solid lines in Fig.~\ref{fig:eP}) this contribution is dominated by bound  states $n<50$.
We reiterate that $n=100$ is sufficient to find a convergent effective cross section, hence even higher bound states  are expected to not alter our results.

For illustration, in Fig.\,\ref{fig:eP}, we include decay rates down to around $10^{-18}\,$GeV for which a significant fraction of mediators have not decayed at $T=1\,$GeV. In this region, the gray shaded areas conservatively bracket the uncertainty in the effective annihilation rate arising from the impact of confinement, by assuming that all mediators that are still present at $T=1$\,GeV either vanish (upper boundary) or fully decay into dark matter particles (lower boundary). However, in this work, we focus on the perturbative regime, $\Gamma_{\tilde q\to q\chi}\ge 10^{-17}$\,GeV, for which we find that the difference between the upper and lower boundary is less than 1\%.

Due to the late decay and large mass difference between the mediator and dark matter, the dark matter momentum distribution for the considered scenario can be significantly harder than the one of cold dark  matter. The resulting free-streaming effect impacts structure formation on small scales probed by Lyman-$\alpha$ forest observations. As shown in Ref.~\cite{Decant:2021mhj}, this results in a lower bound on the dark matter mass,
\begin{equation}\label{eq: Lyman Alpha}
\frac{m_\chi}{\text{keV}} > 3.8\times x_\text{decay} \left(\frac{106.75}{g_{*S}(x_\text{decay})}\right)^{\!1/3\,} \,,
\end{equation}
that can easily reach into the GeV range. Here, \mbox{$x_\text{decay}=(\Gamma_{\tilde q\to q\chi}/H(m_{\tilde q}))^{-1/2}$} is the temperature parameter at which the decay becomes efficient, \mbox{$\Gamma_{\tilde q\to q\chi}=H(m_{\tilde q}/x_\text{decay})$}, \cf~upper axis labeling in Fig.\,\ref{fig:eP}. The formula assumes $\Omega_\chi h^2=0.12$ and $m_q\ll m_{\tilde q}$.

In Fig.\,\ref{fig:eP}, we display the corresponding 95\%~C.L. exclusion as a red shaded area. Interestingly, excited bound state effects have a significant impact on the implications of this constraint. While with Sommerfeld effect (and $n=1$ bound state) only, a decay rate of around $10^{-15} \; (4\times 10^{-16})$\,GeV would be excluded, the inclusion of excited bound states reveal that the entire region with \mbox{$\Gamma_{\tilde q\to q\chi}\ge 10^{-17}$}\,GeV remains unchallenged by the considered Lyman-$\alpha$ constraint. 

So far, we have focused on the benchmark mass \mbox{$m_{\tilde q}=4\times 10^6$\,GeV}\@. For smaller masses, the effective cross section becomes larger and the mediator abundance decreases. Therefore, for a given $x_\text{decay}$, the dark matter mass that leads to $\Omega h^2=0.12$ increases and the \mbox{Lyman-$\alpha$} bound become less constraining. Accordingly,  we find that current Lyman-$\alpha$ forest constraints do not challenge the superWIMP scenario for $m_{\tilde q}<4\times 10^6$\,GeV, if we restrict ourselves to decays within the  perturbative regime of couplings, \ie~decays that take place before the QCD phase transition, $x_\text{decay}< m_{\tilde q}/1\,$GeV.
However, as the so-defined maximal $x_\text{decay}$ becomes smaller with smaller $m_{\tilde q}$, highly excited states become less relevant somewhat diminishing the large effect of bound states towards small $m_{\tilde q}$ in our setup.

For masses larger than $4\times 10^6$\,GeV, the cosmologically viable dark matter mass decreases and Lyman-$\alpha$ constraints become more restrictive. In particular, starting from masses somewhat above $m_{\tilde q}=4\times 10^6$\,GeV, they impose an upper bound on $x_\text{decay}$ that is more restrictive than the above-mentioned perturbativity condition. As a consequence of the tightening Lyman-$\alpha$ constraints on the  allowed range of mediator decay rates, higher excitations become less important also towards larger masses. Eventually, for $m_{\tilde q}>4\times10^8$\,GeV, the entire superWIMP region is excluded (assuming thermalization of the mediator). 

%===================================================================
\section{Conclusion}\label{sec:conclusion}
%===================================================================

In this work, we studied the impact of excited bound states on dark matter production. We found that they can be highly relevant, focusing especially on setups where unbroken Abelian and/or non-Abelian gauge interactions are responsible for the bound state dynamics.
Considering bound state formation/ionization and \mbox{(de-)excitation} processes described by dipole transitions, we developed an efficient way to compute their rates numerically. It allows us to take into account excitations up to a principal quantum number $n=100$ ($n=1000$) in the presence (absence) of transitions among them involving more than 5000 individual BSF and $\mathcal{O}(10^6)$ transitions rates.

With this numerical tool at our disposal, we investigated several theoretical and phenomenological questions. First, considering an Abelian gauge theory like dark QED, we compared the summed BSF cross section to the well-known Kramer's logarithm, confirming its approximate behavior towards small $v$ that increases faster than $\propto 1/v$. However,  this behavior results from a summation over different \emph{initial} state partial waves. We checked that each initial angular momentum contribution is compatible with partial-wave unitarity for all velocities, and at sufficiently weak coupling. In contrast, we found that the total BSF cross section in non-Abelian gauge theories generally does violate partial-wave unitarity bounds even for perturbatively small coupling. While a closer investigation of this feature is left for future work, we focused on the phenomenologically relevant regime of velocities in our subsequent results for which unitarity bounds are satisfied. 

We exemplified our results for two particle states with constituents transforming as $3$ and $\bar{3}$ under $SU(3)$. Due to the repulsive potential of the adjoint scattering state, in radiative BSF via gluon emission, each individual bound state cross section only contributes around a characteristic velocity. This renders very high excitations to be the dominant contribution to the thermally averaged effective cross section at very low temperatures. 
Consequently, we investigated the important question whether the effective cross section increases slower or faster than $x=m/T$, in which case the particle's number density would freeze-out or would continue to deplete. 

For dark QED, we found a scaling $\langle\sigma v\rangle_{\text{eff}}\propto x^{0.6}$ towards large $x$. For dark QCD, $\langle\sigma v\rangle_{\text{eff}}$ grows slightly slower (faster) than $\propto x$, if we exclude (include) the effect of a running coupling (assuming a negative beta function). However, the dependence of this qualitative difference on the effects due to the running coupling is only present in (dark) QCD, \ie~in the absence of transitions among the states. When color and electric charge is combined, then singlet-singlet transitions are possible and significantly enhance the effective cross section towards small $x$ where higher $n$ dominate, and steepen the effective cross section beyond $\propto x$ regardless of running effects.

Finally, we studied such a case in detail by considering a scenario where an electrically and color charged mediator -- which accompanies dark matter -- is subject to the formation of bound states. Such a setup is realized in so-called $t$-channel mediator models. We focused on the very weak coupling regime for which the dark matter density can be generated through a late decaying mediator particle, \ie~by the superWIMP mechanism. The commonly adopted assumption within this paradigm is that mediator freeze-out and decays can be considered independently, thereby rendering the resulting dark matter density to be independent of the  the mediator lifetime (and therewith the dark matter coupling it depends on).

Here we found that considering excited bound state effects, this picture has to be revised. Due to an interplay of QCD bound states and transitions mediated by QED, the resulting effective cross section grows significantly faster than $\propto x$, \ie~features a strongly super critical scaling, thereby retaining a  sizeable depletion of mediator  particles throughout its entire presence before it decays. Therefore, the dark matter relic density \emph{does} depend on the mediator lifetime. In fact, we found that -- restricting ourselves to decays that happen well within the unconfined phase -- the yield reduction due to bound state effects can amount to an order of magnitude with respect to the result including Sommerfeld enhanced annihilation only. This has important consequences for the cosmologically  viable parameter space. For a given  mediator mass, bound state effects shift the dark matter mass required to match $\Omega h^2$ towards larger values by up to an order of magnitude, thereby weakening constraints from structure formation through Lyman-$\alpha$ observations.

While we are choosing dark matter physics as an application of the considered bound state effects, we note that they could also play an important role in the context of baryogenesis. For instance, according to~\cite{Heeck:2023lxl}, the late decay of a color-charged scalar can generate the matter-antimatter asymmetry. A significant enhancement of the effective cross section due to bound states could open up parts of the parameter space otherwise excluded by strong bounds from $N_\text{eff}$. In addition, an analysis of phenomenological consequences within dark sector models, for which dark matter is charged under an unbroken dark gauge symmetry, would be interesting, also in view of phase transitions and associated gravitational wave signatures~\cite{Asadi:2021pwo,Asadi:2022vkc}. On the theoretical side, our results motivate an investigation of unitarization of bound state formation processes mediated by unbroken non-Abelian gauge interactions within the regime of perturbatively small couplings.

%===================================================================
\section*{Acknowledgments}
%===================================================================

We are thankful to Martin Beneke, Florian Herren, and Gramos Qerimi
for discussions. This work was supported by the DFG Collaborative Research Institution Neutrinos and Dark Matter in Astro- and Particle Physics (SFB 1258) and the Collaborative Research Center TRR 257. J.H.~acknowledges support by the Alexander von Humboldt foundation via the Feodor Lynen Research Fellowship for Experienced Researchers and Feodor Lynen Return Fellowship.

\begin{appendix}

%===================================================================
\section{Evaluation of large \emph{n} dipole transitions}\label{sec:bsfapp}
%===================================================================

The evaluation of ultra-soft transitions such as bound-state formation and de-excitation for large $n$ is a demanding task. Here, we briefly summarize the derivation of the used expressions that enable a numerically stable and sufficiently fast computation of the scattering-to-bound and bound-to-bound dipole transitions up to about $n=1000$ and $n=100$, respectively. A \textsc{Mathematica} notebook containing the implemented expressions is available on request.

%-------------------------------------------------------------------
\subsection{Scattering-to-bound}\label{sec:stb}
%-------------------------------------------------------------------
We turn now to the evaluation of the scattering-to-bound dipole transitions matrix elements. The angular part can be performed as~\cite{Garny:2021qsr}:
\begin{align}\label{eq:BSFMatrixelement}
\sum_{m}|\bra{\psi_{n \ell m}} \mathbf{r} \ket{\psi_{\mathbf{p}}^\prime}|^2 = \frac{4 \pi}{|\mathbf{p}|^5} \sum_{\ell^\prime}(\ell^\prime \delta_{\ell^\prime,\ell+1} + \ell \delta_{\ell,\ell^\prime+1}) |I_R|^2,
\end{align}
where we have $|{\bf p} |=mv/2$. Our starting point for the evaluation of the radial integral part is (A.13) in Ref.~\cite{Garny:2021qsr}: 
\begin{align}
  I_{R,\text{BSF}} =& \frac{2\zb^{3/2}(\ell+\ellp+3)!}{n^2 \sqrt{(n-\ell-1)!(n+\ell)!}}\frac{2^{\ellp} e^{\pi\zs/2}}{|\Gamma(1+\ellp-i\zs)|}\times \nonumber \\
  & \left(\frac{2\zb}{n}\right)^\ell 
  \left(\frac{d}{dt}\right)^{n-\ell-1}\frac{1}{(1-t)^{2\ell+2}} \times \nonumber \\
  & \int_0^1 ds \frac{s^{\ellp-i\zs}(1-s)^{\ellp +i\zs}}{\left(\frac{\zb}{n}\frac{1+t}{1-t}+i(2s-1)\right)^{\ell+\ellp+4}}\Big|_{t=0}\,,\label{eq:irgarny}
\end{align}
where $\zeta_b=\alpha^{\text{eff}}_b/v$ and $\zeta_s=\alpha^{\text{eff}}_s/v$.
To arrive at this expression, the representation of the bound-state wave function in terms of associated Laguerre polynomials and their generating function was used while for the scattering state the Hypergeometric function ${}_1F_1$ in terms of the integral representation was used. In this work, we directly evaluate the integral first as
\begin{align}
&\int_0^1 ds \frac{s^{\ellp-a}(1-s)^{\ellp +a}}{\left(-i b+i(2s-1)\right)^{\ell+\ellp+4}} = \nonumber \\
& (-i)^{-\ell-\ellp} \, \frac{ \Gamma(1-a+\ellp) \Gamma(1+a+\ellp)}{(1+b)^{4 +\ell + \ellp} \Gamma(2 (1+\ellp))} \, \times \nonumber \\ & {}_2F_1\left(1-a+\ellp, 4+\ell +\ellp, 2 (1+\ellp), \frac{2}{1+b}\right)\,,
\label{eq:sInt}
\end{align}
where $a =i \zs, b = i \frac{\zb}{n} \frac{1+t}{1-t}$. Applying the $t$ derivatives in Eq.~(\ref{eq:irgarny}) to the Hypergeometric function ${}_2F_1$, allows us to obtain the following recursive relations:
\begin{align}
& |I_{R,\text{BSF}}^{\ellp = \ell +1}| = 2 n^3 \sqrt{(\ell^2 + \zs^2) ((\ell +1 )^2 + \zs^2)} P(n,\ell) \times \nonumber \\
& \Big|((2+\ell) \zbt + \zs) R_{n,\ell}(n-\ell-5) \nonumber \\
& - 2 ((2+\ell) \zbt + 2 \zs) R_{n,\ell}(n-\ell-4)  \nonumber \\ 
&+ 6 \zs R_{n,\ell}(n-\ell-3)   \nonumber \\ 
&+ 2 ((2+\ell) \zbt -2 \zs) R_{n,\ell}(n-\ell-2) \nonumber \\ 
&+ (\zs - (2+\ell) \zbt) R_{n,\ell}(n-\ell-1) \Big| \; ,
\end{align}
and
\begin{align}
& |I_{R,\text{BSF}}^{\ellp = \ell  - 1} |= n^3 P(n,\ell) \times \nonumber \\
& \Big|- \big[(\ell (1 + \ell) \zbt (-3 + (1+2\ell)) \zbt^2) \nonumber \\
&+ (-1-3\ell +3(1+\ell) (1+2\ell) \zbt^2) \zs  \nonumber\\ 
&+ 6 (1+\ell) \zbt \zs^2 + 2 \zs^3\big] R_{n,\ell}(n-\ell-5) \nonumber \\
& -2 \big[ \ell (1+\ell) \zbt (3 + (1+2\ell) \zbt^2) + 2 \zs + 6 \ell \zs \nonumber \\
&- 6 (1+\ell) \zbt \zs^2 -4 \zs^3 \big] R_{n,\ell}(n-\ell-4) \nonumber \\
& + \big[6 \zs (1 + 3 \ell + (1+\ell) (1+2\ell) \zbt^2 - 2 \zs^2)\big]R_{n,\ell}(n-\ell-3) \nonumber \\
& + 2 \big[\ell (1+\ell) \zbt (3 + (1+2\ell) \zbt^2) - 2 (1+3 \ell) \zs \nonumber \\ 
&- 6 (1+\ell) \zbt \zs^2 + 4 \zs^3\big] R_{n,\ell}(n-\ell-2)\nonumber \\
&+\big[\ell (1+\ell) \zbt (-3 + (1+2\ell) \zbt^2) + \zs + 3 \ell \zs \nonumber \\
&- 3 (1+\ell) (1+2\ell)\zbt^2 \zs \nonumber \\
&+ 6 (1+\ell) \zbt \zs^2 - 2 \zs^3\big] R_{n,\ell}(n-\ell-1)\Big| \; ,
\end{align}
where $\zbt = \zb / n$. The common prefactor is 
\begin{align}
P(n,\ell) &= \sqrt{\frac{(n-\ell-1)!}{(n+\ell)!}}  \frac{2^{2+2\ell} \zbt^{\ell + \frac{3}{2}}}{n^{\frac{7}{2}} \zs (1+\zbt^2)^{3+\ell}} \times \nonumber \\ &  \sqrt{\frac{2 \pi \zs}{1-e^{-2 \pi \zs}}} e^{-2 \zs \text{arccot}(\zbt)} \sqrt{ \prod_{j = 0}^{\ell -1} \left( j^2 + \zs^2 \right) }
\label{eq:prefactor}
\end{align}
and the common recursion is given by
\begin{align}
R_{n,\ell}(x) = \left\{
\begin{array}{lr}
0 & x < 0 \\
1 & x = 0 \\
\frac{-2 (3+\ell) (\zbt^2-1) + 4 \zbt \zs}{1+ \zbt^2} & x = 1 \\
\frac{ 2 (2+\ell+x) (1-\zbt^2) + 4 \zbt \zs}{x (1+\zbt^2)} R_{n,\ell}(x-1) & \\
\;\;- \,\frac{(4+2\ell+x)}{x}  R_{n,\ell}(x-2)  & {\rm else}\,.
\end{array}
\right. 
\label{eq:recursion}
\end{align}

The expressions allow for a fast generation of the matrix elements entering the BSF cross section, also for large $n,\ell$. They have been cross-checked against the expressions in Ref.~\cite{Garny:2021qsr} for $n \leq 10$ and $\ell\leq n-1$ for the QCD case, as well as for the simpler QED limit $\zeta_s = \zeta_b = \zeta$. To achieve the results presented in this work up to $n=1000$ we split the prefactor (\ref{eq:prefactor}) into two multiplicative pieces to avoid numerical underflow. 

%-------------------------------------------------------------------
\subsection{Bound-to-bound}\label{sec:btb}
%-------------------------------------------------------------------
We turn now to the evaluation of the bound-to-bound dipole transition matrix elements. The angular part can be performed as~\cite{Garny:2021qsr}:
\begin{align}
\sum_{m,m^\prime}|\bra{\psi_{n \ell m}} \mathbf{r} \ket{\psi_{n^\prime \ell^\prime m^\prime}^\prime}|^2 &= \nonumber \\
(\ell^\prime \delta_{\ell^\prime,\ell+1} & + \ell \delta_{\ell,\ell^\prime+1}) |I_{R,\text{trans}}|^2\,,
\label{eq:btbMatrixElement}
\end{align}
leaving only the radial integral over the initial and final bound state wave-functions. For convenience, we define $|\bra{\psi_{n\ell}} \mathbf{r} \ket{\psi_{n^\prime \ell^\prime}}|^2$ as the squared matrix element average over $m, m'$. We start our evaluation, by representing both wave functions in terms of the Hypergeometric functions ${}_1F_1$.\footnote{
Another evaluation can be made by representing the wave functions of the bound states in terms of the associated Laguerre polynomials:
\begin{align}
L_{n}^{\alpha}(x) = \sum_{i = 0}^{n} (-1)^i \left(\begin{array}{c}
n + \alpha \\
n - i
\end{array} \right) \frac{x^i}{i!}.
\label{eq:Lag}
\end{align}
Performing the radial integral in this representation leads to
\begin{align}
&I_{R,{\rm trans}} = N_{n\ell}(\kappa) N_{\np \ellp}(\kp)(\kt + \kpt)^{-4-\ell-\ellp } \times \\ & \sum_{k = 0}^{n-\ell-1} \sum_{k^\prime = 0 }^{\np - \ellp - 1} \left(\begin{array}{c}
n+\ell \\
n-\ell-k-1
\end{array}\right) \left(\begin{array}{c}
\np+ \ellp \\
\np-\ellp-k^\prime-1
\end{array}\right) \times \nonumber \\
& \frac{(-2)^{k + k^\prime}   \Gamma(4+\ell+\ellp + k + k^\prime)}{k! k^\prime! }  \frac{\kt^k (\kpt)^{k^\prime}}{(\kt + \kpt)^{k + k^\prime}}\,.
\end{align}
With the help of \textsc{Mathematica}, both sums can be performed, resulting in similar recursions as we have obtained for the scattering-to-bound case. In practice, however, it turns out that these bound-to-bound recursions are less numerically stable within standard digit precision than what we have obtained by using the Hypergeometric functions in Eqs.~(\ref{eq:trlp1}),(\ref{eq:trlm1}) and (\ref{eq:subst}). 
} 
As pointed out recently in Ref.~\cite{Biondini:2023zcz}, one can perform the radial integration and obtain closed expressions in terms of Hypergeometric functions by following Ref.~\cite{GordonZurBD}. Adopting this procedure, we obtain for the general case
\begin{align}
&|I_{R,\text{trans}}^{\ellp = \ell +1}|= N_{n,\ell}(\Tilde{\kappa})N_{n^\prime,\ell+1}(\Tilde{\kappa}^\prime) \times \label{eq:trlp1}\\ & 2^{-1} (\ell+1)(2\ell+3)\Gamma(2\ell+2) (1-z)^{\frac{n^\prime -n}{2}}\frac{z^{\ell+1}}{(\Tilde{\kappa}^\prime)^2\Tilde{\kappa}}\times \nonumber \\
&\bigg| \frac{\Tilde{\kappa}^\prime - \Tilde{\kappa}n +\Tilde{\kappa}^\prime n^\prime }{(\Tilde{\kappa}+\Tilde{\kappa}^\prime)^2}  \prescript{}{2}{F}{_1}(-n+\ell+1;n^\prime + \ell+2;2\ell+2;z) \nonumber \\
&+ 2 \frac{\Tilde{\kappa}n -\Tilde{\kappa}^\prime n^\prime }{(\Tilde{\kappa}^2-\Tilde{\kappa}^{\prime 2})}  \prescript{}{2}{F}{_1}(-n+\ell+1;n^\prime+\ell+1;2\ell+2;z)   \nonumber \\
&- \frac{\Tilde{\kappa}^\prime + \Tilde{\kappa}n -\Tilde{\kappa}^\prime n^\prime }{(\Tilde{\kappa}-\Tilde{\kappa}^\prime)^2}  \prescript{}{2}{F}{_1}(-n+\ell+1;n^\prime+\ell;2\ell+2;z)  \nonumber \bigg|\,,
\end{align}
and
\begin{align}
&|I_{R,\text{trans}}^{\ellp = \ell - 1}|=N_{n,\ell}(\Tilde{\kappa})N_{n^\prime,\ell-1}(\Tilde{\kappa}^\prime) \times \label{eq:trlm1}\\ & 2^{-1} \ell(2\ell+1)\Gamma(2\ell) (1-z)^{\frac{n^\prime -n}{2}} \frac{z^\ell}{\Tilde{\kappa}^2 \Tilde{\kappa}^\prime}\times \nonumber \\
&\bigg| -\frac{\Tilde{\kappa}+\Tilde{\kappa}n -\Tilde{\kappa}^\prime n^\prime }{(\Tilde{\kappa}-\Tilde{\kappa}^\prime)^2}  \prescript{}{2}{F}{_1}(-n+\ell-1;n^\prime+\ell;2\ell;z) \nonumber \\
&- 2 \frac{\Tilde{\kappa}n -\Tilde{\kappa}^\prime n^\prime }{(\Tilde{\kappa}^2-\Tilde{\kappa}^{\prime 2})}  \prescript{}{2}{F}{_1}(-n+\ell;n^\prime+\ell;2\ell;z)   \nonumber \\
&+ \frac{\Tilde{\kappa} - \Tilde{\kappa}n +\Tilde{\kappa}^\prime n^\prime }{(\Tilde{\kappa}+\Tilde{\kappa}^\prime)^2}  \prescript{}{2}{F}{_1}(-n+\ell+1;n^\prime+\ell;2\ell;z)  \nonumber \bigg|\,,
\end{align}
with $z=\frac{4  \Tilde{\kappa} \Tilde{\kappa}^\prime }{(\Tilde{\kappa}+\Tilde{\kappa}^\prime)^2}<1$ and $\Tilde{\kappa}= \alpha \mu / n$, $\Tilde{\kappa}^\prime= \alpha^\prime \mu^\prime / n^\prime$ and $\Tilde{\kappa}\neq \Tilde{\kappa}^\prime$. For the special case $\alpha^\prime \mu^\prime =\alpha \mu$ (particularly important for QED), see Ref.~\cite{Biondini:2023zcz, GordonZurBD}.  The normalization is given by:
\begin{align}
N_{n,\ell}(\Tilde{\kappa})= \frac{\Tilde{\kappa}^{3/2}}{\sqrt{n}} \frac{2}{(2\ell+1)!} \left( \frac{(n+\ell)!}{(n-\ell-1)!}\right)^{1/2}\,.
\end{align}

For large initial state principle quantum numbers $n=\mathcal{O}(10)$, the numerical evaluation of the Hypergeometric functions can be problematic for transitions where $z \rightarrow 1 $ ($n^\prime \rightarrow n$). To improve the numerical stability in this regime, we use a lengthy expression for the Hypergeometric function expanded around $1-z$. For the special arguments of the Hypergeometric functions as given above, we can further simplify the expression and finally obtain to all orders in $1-z$:
\begin{align}
&\prescript{}{2}{F}{_1}(a;b;c;z) \rightarrow (-1)^{-a} (-1 + z)^{-a - b + c} \times \label{eq:subst} \\ & \frac{\Gamma(1-a) \Gamma(c)}{
 \Gamma(b) \Gamma(1 \!-\! a\! -\! b + c)}
  \prescript{}{2}{F}{_1}(c - a; c - b; c - a - b + 1; 1 - z) \,.\nonumber
\end{align}
Notice that this substitution is an identity for all sets of $\{a;b;c;z\}$ arguments given above. In practice, we use the substitution for $z>0.7$, which allows us to obtain stable numerical results for all bound-to-bound transitions with $n \leq 100$.
Increasing the number of digits for $z$ (\eg~via \texttt{MaxExtraPrecision} in \textsc{Mathematica}) allows for even larger $n$ values and also to check the stability, with the cost of loosing efficiency.

%===================================================================
\section{Cross sections and rates}\label{sec:AppRates}
%===================================================================

%-------------------------------------------------------------------
\subsection{Thermal average and Milne relations}\label{sec:ThAvMilne}
%-------------------------------------------------------------------

In the non-relativistic limit, the thermal average  of the bound state formation cross section can be written as
\begin{equation}
  %\langle\sigma_{\BSF,i}v\rangle =
  \avb{\sigma v}_{i} =
  \left(\frac{\mu}{2\pi T}\right)^{3/2} \! \int \!\diff^3 v \, \E^ {-\frac{\mu v^2}{2T}}  \left[1+f(\Delta  E)\right]\, 
  %\sigma_{\BSF,i} v 
  (\sigma v)_i\,,
  \label{eq:sigmaBSFav}
\end{equation}
where $f(\Delta  E)=1/(\E^{\Delta  E/T}-1)$. We note that the thermal averages in non-Abelian gauge theories need to be evaluated carefully due to oscillatory features.\footnote{For example, for effective coupling $\alpha_s^\text{eff} < 0 < \alpha_b^\text{eff}$ relevant in $SU(N_c)$, $(\sigma v)_{n\ell}$ features $n-\ell-1$ local minima in its velocity dependence. Fewer local minima arise when $0<\alpha_s^\text{eff} < \alpha_b^\text{eff}$.}

As usual, the cross section includes an average of initial state degrees of freedom and a sum over final state degrees of freedom. For the case of bound states with Fermionic spin-$1/2$ constituents, one has to distinguish between bound states in a spin-singlet or spin-triplet configuration. The main difference between the two cases occurs for the bound state decay rate, see below. Since the electric dipole interaction is spin-independent, bound-to-bound transitions do not change the spin within our approximations, \ie~transitions occur  exclusively among spin-singlet states or spin-triplet states, respectively. The transition rates are identical in both cases. Furthermore, for scattering-to-bound processes, the contribution to the BSF cross section for formation of any single spin degree of freedom of the bound state is the same. Therefore, we account for the formation of spin-singlet or spin-triplet bound states by the prefactor $\xi=1/4$ or $\xi=3/4$, respectively, in Eqs.~\eqref{eq:sigBSFdarkQED} and~\eqref{eq:sigBSFdarkQCD}. Furthermore, the BSF cross sections also apply for bound states composed of charged scalars, with $\xi=1$. The reason is that the factor $1/4=1/2\times 1/2$ from averaging over initial state spin and the factor $4=3+1$ from summing over all final state spin configurations cancel out in the Fermionic case.

For computational efficiency, 
we use the following Milne relations based on detailed balance 
%of the micro physics 
to obtain the inverse processes of BSF, \ie~ionization, and bound state de-excitation, \ie~excitation. Assuming leading order non-relativistic expansions for the equilibrium yields $Y_{j}^\text{eq}$ (the combined particle and anti-particle yield) and $Y_{\mathcal{B}_i}^\text{eq}$, the ionization rate can be expressed as
\be \label{eq:Milne}
  \Gamma_\text{ion}^i = 
  %\frac{1}{\xi}
  \frac{s}{4} \, \frac{ (Y_{j}^{\eq})^2 }{ Y_{{\cal B}_i}^{\eq} } \avb{\sigma v}_i \,.
\ee
Note that for bound states with Fermionic constituents, it holds for both spin-singlet and spin-triplet states when using the appropriate BSF cross section as described above and accounting for the factors of spin degrees of freedom contained in the equilibrium yields.

The transition rates among the set of bound states are similarly related via
\be \label{eq:transMilne}
  \Gamma_\text{trans}^{i\to j} = \Gamma_\text{trans}^{j\to i} \frac{ Y_{{\cal B}_j}^\eq }{ Y_{{\cal B}_i}^\eq }\,.
\ee

%-------------------------------------------------------------------
\subsection{Dark QED}\label{sec:AppDarkQED}
%-------------------------------------------------------------------

In the absence of light Fermions the coupling is not running within dark QED   and, hence, the effective coupling strengths are identical for bound and scattering states, \ie~$\alpha_s^\text{eff}=\alpha_b^\text{eff}=\alpha$ here. 
The annihilation cross section into a pair of massless dark photons, including Sommerfeld enhancement, is given by~\cite{doi:10.1002/andp.19314030302,Sakharov:1948yq}
\begin{align}
(\sigma v)_\text{ann}= \frac{\pi \alpha^2}{m^2}\, S_0\left(\frac{\alpha}{v}\right)\,,\label{eq:annQED}
\end{align}
where 
\begin{equation}\label{eq:SEdef}
    S_0(\zeta)\equiv\frac{2\pi\zeta }{1-e^{-2\pi\zeta}} \,.
\end{equation}

The BSF cross section, which includes the summation over the degenerate magnetic quantum number of the final state $n\ell$ and also considers all possible initial angular momenta, is in the case of $U(1)$ interactions given by 
\begin{align}
    \label{eq:sigBSFdarkQED}
    \left(\sigma v\right)_{n\ell} = \xi\frac{\pi \,\alpha^2 }{m^2}\frac{2^9}{3}
    \, S_\text{BSF}(n,\ell,\frac{\alpha}{n v},\frac{\alpha}{v}) \,,
\end{align}
where $\xi$ accounts for spin factors as defined in Sec.\,~\ref{sec:ThAvMilne}, and
\begin{align}
    S_\text{BSF}(n,\ell,\tilde{ \zeta}_b,&\zeta_s) \equiv \,\frac{1}{2^6n \tilde{\zeta}_b}\left(1+{\tilde{\zeta}_b}^2\right)^3 \times \notag \\ 
    &\left[ (\ell+1)|I_R^{\ell'=\ell+1}|^2+\ell|I_R^{\ell'=\ell-1}|^2\right].
    \label{eq:SBSF}
\end{align}

$S_\text{BSF}$ can be evaluated numerically in an efficient way by using the radial integral formulas laid out in App.~\ref{sec:stb}. Further analytic simplifications due to the absence of scale running of the coupling strength and the identical initial and final state gauge representations are possible, though not used in our numerical evaluation.

For spin-singlet bound states, we adopt the decay rate into two dark photons for the s-wave states from Ref.~\cite{pirenne1946proper,doi:10.1111/j.1749-6632.1946.tb31764.x}: % 
\begin{align}
\Gamma_\text{dec,\,QED}^{n\ell\to \gamma \gamma} &= \delta_{\ell,0}\,\frac{m\,\alpha^5}{2n^3} \,.
\end{align}
For spin-triplet bound states we adopt the decay rate into three dark photons for the s-wave states from Ref.~\cite{1949PhRv...75.1696O}: 
\begin{align}
\Gamma_\text{dec,\,QED}^{n\ell\to 3 \gamma} &=  \frac{4(\pi^2-9)\alpha}{9 \pi} \times \Gamma_\text{dec,\,QED}^{n\ell\to \gamma \gamma}  \,.
\end{align}

Lastly, the de-excitation rate relates to the dipole matrix element via
\bea
    \Gamma_\text{trans,\,QED}^{n'\ell'\to n\ell} &=& \frac{4 \alpha}{3} (2\ell+1)\left(\frac{m\alpha^2}{4}\left|\frac{1}{n^2}-\frac{1}{n'^2}\right|\right)^3 \nn\\
    && {} \times |\bra{\psi_{n^\prime \ell^\prime}} \mathbf{r} \ket{\psi_{n\ell}}|^2 ,
\eea
according to App.~\ref{sec:btb}.

%-------------------------------------------------------------------
\subsection{Dark QCD}\label{sec:AppDarkQCD}
%-------------------------------------------------------------------

In Yang-Mills-theories, gauge Boson self interactions give rise to running of the coupling strength, even in the absence of light Fermions. For our numerical benchmark, we defined the value $\alpha(m)\,\equiv\,0.025$ and employ one-loop running. Using this choice, the non-perturbative regime $\alpha(m/x)=1$ starts at $x\approx 4\times 10^9$, which holds for any mass since $m$ must drop out of dimensionless expressions, being the only mass scale in the theory.
The heavy Fermions are from the fundamental representation of $SU(3)$ and thus can form singlet (${\bf1}$) and octet (${\bf8}$) two particle states. We evaluate the wave functions of the initial scattering (bound) states, $s$ ($b$), at their respective (Bohr-) momentum scale. The effective couplings are thus given by
\begin{align}\label{eq:alphaBeff}
    \alpha_b^\text{eff} &\equiv 
    % \frac{4}{3}
    C_F\,\alpha\left(\frac{m}{2}\frac{\alpha_b^\text{eff}}{n}\right)\,,\\
    \alpha_s^\text{eff} &\equiv 
    %-\frac{1}{6}
    \frac{2C_F-C_A}{2}\,\alpha\left(\frac{m}{2}v\right)\,, 
\end{align}
where $C_F=4/3$ and $C_A=3$. Eq.~(\ref{eq:alphaBeff}) is an implicit definition easily solved for, either numerically or analytically order by order. 
The annihilation into two gluons is possible from singlet or octet scattering states and takes the form
\begin{align}\label{eq:Annh} 
(\sigma v)_\text{ann}= \frac{7}{27}\frac{\pi \alpha(2m)^2}{m^2} \left( \frac27 S_0^{[\bf{1}]} + \frac57 S_0^{[\bf{8}]}\right) \,.
\end{align}
The Sommerfeld factors are given by
\begin{align}
S^{[\bf{1}]}_0 &= \frac{\alpha(mv/2)}{v}\frac{2\pi C_F}{ 1 - e^{-2\pi C_F \alpha(m\frac v2)/v}} \,, 
\\
S^{[\bf{8}]}_0 &= \frac{ \alpha(mv/2)}{v}\frac{2\pi(C_F-C_A/2) }{ 1-e^{-2\pi (C_F-C_A/2)\alpha(mv/2)/v}} \,.
\end{align}

The BSF cross section in a general $SU(N_c)$ theory takes the form~\cite{Garny:2021qsr} 
\begin{align}\label{eq:sigBSFdarkQCD}
    % \avb{\sigma v}_{n\ell} = 
    \left(\sigma v\right)_{n\ell} = 
    \xi\frac{\pi \,\alpha_b^\text{eff}\, \alpha_\text{BSF}}{m^2}\frac{2^9C_F}{3N_c^2}
    \, S_\text{BSF}(n,\ell,\frac{\alpha_b^\text{eff}}{n v},\frac{\alpha_s^\text{eff}}{v}),    
\end{align}
with $\alpha_\text{BSF} = \alpha\left(mv^2/4 + E_{{\cal B}_{n\ell}}\right)$ and $S_\text{BSF}$ defined in Eq.~(\ref{eq:SBSF}). Here $\xi$ accounts for spin factors as defined in Sec.\,~\ref{sec:ThAvMilne}. The spin-singlet bound state decay rate into two gluons is given by~\cite{Harz:2018csl,Garny:2021qsr,Biondini:2023zcz}
\begin{align}
\Gamma_\text{dec,\,QCD}^{n\ell\to gg} &= \delta_{\ell,0}\,\frac{m\,C_F}{4n^3} \alpha(m)^2 
 \left(\alpha_b^\text{eff}\right)^3 \,.
\end{align}
We neglect spin-triplet bound states for dark QCD.

Dark QCD automatically corresponds to the limiting case of no transitions, \ie~$\Gamma_\text{trans}^{i\to j} = 0$, therefore the effective cross section Eq.~(\ref{eq:effgeneral}) simplifies to~\cite{Binder:2021vfo,Garny:2021qsr}

\be
  \avb{ \sigma v }_\text{eff} \to \avb{ \sigma  v }_\text{ann}+\sum_{i} \avb{\sigma v}_i \,
  \frac{ \Gamma^{i}_\text{dec} }{ \Gamma^{i}_\text{ion} + \Gamma^{i}_\text{dec} } \,.
\ee
This result can be seen as a straightforward generalization of the single bound state case~\cite{Ellis:2015vaa}, extended to a sum over individual bound states that do not impact each other. 

%-------------------------------------------------------------------
\subsection{superWIMP scenario}\label{sec:AppSWIMP}
%-------------------------------------------------------------------

The expressions for BSF of a colored $t$-channel scalar mediator is identical to that for dark QCD in Eq.~\eqref{eq:sigBSFdarkQCD}, now using $m=m_{\tilde{q}}$ and the SM strong coupling strength $\alpha_s$  to 5-loop accuracy as implemented in \textsc{RunDec}~3~\cite{Herren:2017osy} in place of $\alpha$, as well as $\xi=1$.  The annihilation cross section reads
\begin{align}
(\sigma v)_\text{ann}= \frac{14}{27}\frac{\pi \alpha(2m_{\tilde q})^2}{m_{\tilde q}^2} \left( \frac27 S_0^{[\bf{1}]} + \frac57 S_0^{[\bf{8}]}\right) \,,
\end{align}
and the decay rate is given by~\cite{Harz:2018csl,Garny:2021qsr}
\begin{align}
\Gamma_\text{dec}^{n\ell\to gg} &= \delta_{\ell,0}\,\frac{m_{\tilde q}\,C_F}{8n^3} \alpha(m_{\tilde q})^2 
 \left(\alpha_b^\text{eff}\right)^3 \,.
\end{align}

Gauge invariance fixes the gauge representations of $\tilde{q}$ to that of the bottom quark in our model, hence the electromagnetic charge is $Q=-1/3$. The addition of electromagnetic interactions leads to transitions  between bound states  in dipole approximation but to no additional relevant decay or BSF channels. 
The transition matrix elements are computed according to App.~\ref{sec:btb} and enter the de-excitation rates as

\begin{align}
    \Gamma_\text{trans,\,SW}^{n'\ell'\to n\ell} = \frac{4 Q^2_{\tilde{q}}\,\alpha_{\text{em}}}{3} (2\ell+1)(\Delta E_{nn'})^3 \,|\bra{\psi_{n^\prime\ell^\prime}^{[\bf1]}} \mathbf{r} \ket{\psi_{n \ell}^{[\bf1]}}|^2\,,
\end{align}
where the fine structure constant is  $\alpha_\text{em}=1/128.9$, and
\begin{equation}
    \Delta E_{nn'} = 
    \frac{m_{\tilde{q}}}{4}\left|
        \frac{\alpha_b^\text{eff}(n)}{n^2}-\frac{\alpha_b^\text{eff}(n')}{n'^2}
    \right|\,.
\end{equation}

%===================================================================
\section{Relic abundance for dark QED}\label{sec:uniQED}
%===================================================================

\begin{figure}[t]
    \centering
    \vspace{4ex}\includegraphics[scale=0.55]{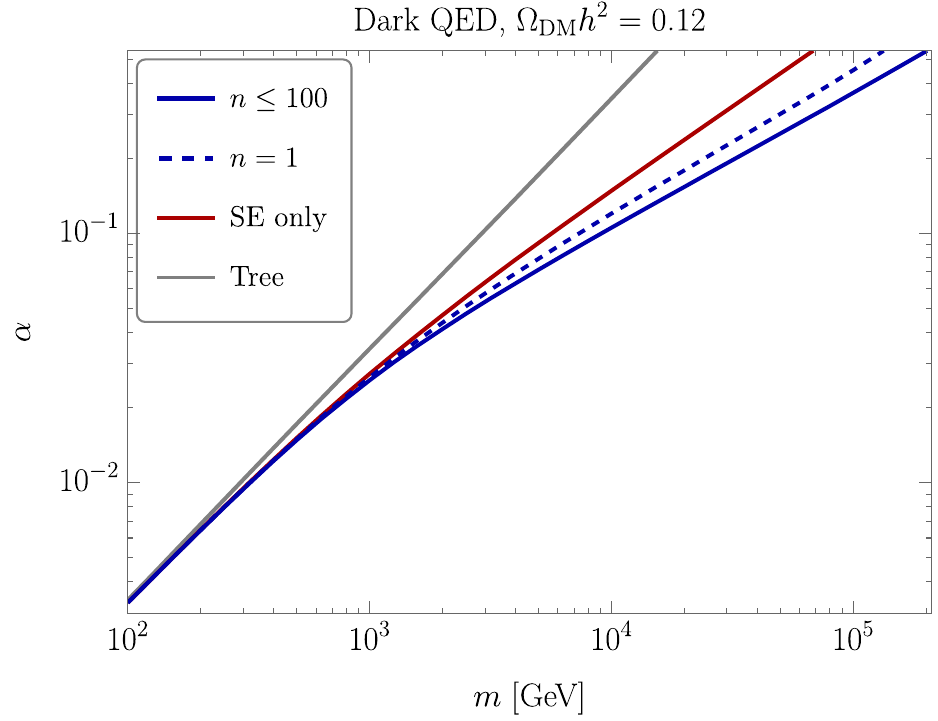}
    \caption{
    \label{fig:uniQED}
    Coupling strength $\alpha$ versus the dark matter mass $m$ shown for various approximations, imposing the right relic abundance constraint. Excited states with $n\leq 100, \ell \leq n-1$ and all dark electric dipole transitions among them (about a million) shown as the blue solid line, are a major correction to the original computation of capture into the ground state only on top of the Sommerfeld effect (blue dashed), \cf~Ref.~\cite{vonHarling:2014kha}.
    }
\end{figure}

Here, we explore the parameter space of dark QED consistent with the relic density measurement. We assume that the dark QED sector is in kinetic equilibrium with the SM heat bath. Solving Eq.~\eqref{eq:simpleBME} for a large number of points in the two-dimensional parameter space of the model, we compute the coupling strength $\alpha$ as a function of the dark matter mass $m$ that provides $\Omega_\chi h^2\simeq 0.12$~\cite{Planck:2018vyg}. 
Figure~\ref{fig:uniQED} displays the respective results under various approximations.  While the known cases of Sommerfeld enhanced annihilation and capture into the ground state further improve the tree-level result, it is visible from the blue solid line that the inclusion of all bound states with $n \leq 100$ and $\ell\leq n-1$ (about 5000 in total), and all possible electric dipole transitions among them (about $10^6$ in total) results in corrections that are worth to us to report.

To arrive at this result, we have included spin-singlet and spin-triplet decay of the s-wave bound states in the effective cross section, such that all curves other than the blue solid line confirm the earlier result in Ref.~\cite{vonHarling:2014kha}. We have observed convergence regarding the number of included bound states and transitions. This completes the discussion within the electric dipole operator picture, in particular, for ultra-soft processes with one dark photon.

\end{appendix}
%%%  Bibliography 
\bibliographystyle{bstyle}
\bibliography{bibliography}{}

\end{document}